\documentclass[a4paper,10pt]{article}
\usepackage{stmaryrd}
\usepackage{amsfonts}
\usepackage{bbm}
\usepackage{amscd}
\usepackage{mathrsfs}
\usepackage{latexsym,amssymb,amsmath,amscd,amscd,amsthm,amsxtra}
\usepackage[dvips]{graphicx}
\usepackage[utf8]{inputenc}
\usepackage[T1]{fontenc}
\usepackage{lmodern}
\usepackage{amssymb}
\usepackage[all]{xy}
\usepackage{nicefrac,mathtools,enumitem}
\usepackage{microtype}

\usepackage{ifpdf}
\ifpdf
 \usepackage[colorlinks,final,backref=page,hyperindex]{hyperref}
\else
 \usepackage[colorlinks,final,backref=page,hyperindex,hypertex]{hyperref}
\fi

\newtheorem{thm}{Theorem}[section]

\newtheorem{cor}[thm]{Corollary}
\newtheorem{pro}[thm]{Proposition}
\newtheorem{ex}[thm]{Example}
\newtheorem{rmk}[thm]{Remark}
\newtheorem{defi}[thm]{Definition}

\newcommand {\emptycomment}[1]{} 

\newcommand{\Symm }{\mathsf{Symm}}
\newcommand{\lon }{\,\rightarrow\,}
\newcommand{\be }{\begin{equation}}
\newcommand{\ee }{\end{equation}}
\newcommand{\ppE}[1]{\left\langle  #1\right\rangle }
\newcommand{\defbe}{\triangleq}
\newcommand{\pf}{\noindent{\bf Proof.}\ }

\newcommand{\h}{\frkh}

\newcommand{\huaS}{\mathcal{S}}
\newcommand{\huaA}{\mathcal{A}}

\newcommand{\huaL}{\mathcal{L}}

\newcommand{\huaE}{\mathcal{E}}

\newcommand{\huaM}{\mathcal{M}}

\newcommand{\huaD}{\mathcal{D}}

\newcommand{\huaK}{\mathcal{K}}

\newcommand{\CWM}{C^{\infty}(M)}

\newcommand{\frka}{\mathfrak a}

\newcommand{\frkd}{\mathfrak d}
\newcommand{\frke}{\mathfrak e}

\newcommand{\frkh}{\mathfrak h}

\newcommand{\frkl}{\mathfrak l}

\newcommand{\frkp}{\mathfrak p}

\newcommand{\frkr}{\mathfrak r}

\newcommand{\frkt}{\mathfrak t}

\newcommand{\frkL}{\mathfrak L}

\newcommand{\frkX}{\mathfrak X}

\def\qed{\hfill ~\vrule height6pt width6pt depth0pt}
\newcommand{\rhowx}{\rho^{\star}}
\newcommand{\half}{\frac{1}{2}}

\newcommand{\ppairingE}[1]{\left ( #1\right )_E}

\newcommand{\Courant}[1]{\left\llbracket  #1\right\rrbracket }
\newcommand{\Poisson}[1]{\{ #1\}}

\newcommand{\br}[1]{   [ \cdot,    \cdot  ]   }

\newcommand{\Lied}{\frkL}
\newcommand{\jet}{\mathfrak{J}}

\newcommand{\jetd}{\mathbbm{d}}
\newcommand{\dev}{\mathfrak{D}}

\newcommand{\id}{\rm{id}}

\newcommand{\g}{\mathfrak g}

\newcommand{\dM}{\mathrm{d}}

\newcommand{\LWX}{\mathsf{CLWX}}
\newcommand{\VB}{\mathsf{VB}}
\newcommand{\Dorfman}[1]{\left  \llbracket  #1\right  \rrbracket }

\newcommand{\Hom}{\mathrm{Hom}}

\newcommand{\gl}{\mathfrak {gl}}

\newcommand{\End}{\mathrm{End}}
\newcommand{\ad}{\mathrm{ad}}

\newcommand{\pr}{\mathrm{pr}}

\newcommand{\sgn}{\mathrm{sgn}}
\newcommand{\Ksgn}{\mathrm{Ksgn}}

\begin{document}
\title{  Categorification of $\VB$-Lie algebroids and  $\VB$-Courant algebroids  }
\author{Yunhe Sheng \\\vspace{2mm}
 Department of Mathematics, Jilin University,
Changchun 130012,  China
\\ Email:  shengyh@jlu.edu.cn
}

\date{}
\maketitle

\vspace{-6mm}
\begin{abstract}
In this paper, first we introduce the notion of a $\VB$-Lie $2$-algebroid, which can be viewed as the categorification of a $\VB$-Lie algebroid. The tangent prolongation of a Lie $2$-algebroid is a $\VB$-Lie $2$-algebroid naturally. We show that after choosing a splitting, there is a one-to-one correspondence between $\VB$-Lie $2$-algebroids and flat superconnections of a Lie 2-algebroid on a 3-term complex of vector bundles.  Then we introduce the notion of a $\VB$-$\LWX$ 2-algebroid, which can be viewed as the categorification of a $\VB$-Courant algebroid. We show that there is a one-to-one correspondence between split Lie 3-algebroids and split $\VB$-$\LWX$ 2-algebroids. The notion of a $\VB$-Lie $2$-bialgebroid is introduced and the double of a $\VB$-Lie $2$-bialgebroid is a  $\VB$-$\LWX$ 2-algebroid. Finally, we introduce the notion of an $E$-$\LWX$ 2-algebroid and show that associated to a $\VB$-$\LWX$ 2-algebroid, there is an $E$-$\LWX$ 2-algebroid structure on the graded fat bundle naturally. By this result, we give a construction of a Lie 3-algebra from a given Lie 3-algebra, which provides interesting examples of Lie 3-algebras including the higher analogue of the string Lie 2-algebra.
\end{abstract}
\tableofcontents

\section{Introduction}

In this paper, we study the categorification of  $\VB$-Lie algebroids and $\VB$-Courant algebroids, and establish the relations between these higher structures and super representations of Lie 2-algebroids,   tangent prolongations of  Lie 2-algebroids,    N-manifolds of degree 3, tangent prolongations of  $\LWX$ 2-algebroids and  higher analogues of the string Lie 2-algebra.

\emptycomment{
The paper is motivated by the following questions:
\begin{itemize}
\item[$\bullet$] Categorified Lie and Courant algebroids are deeply studied recently. What are the categorified structures of a $\VB$-Lie and a $\VB$-Courant algebroid?
  \item[$\bullet$]   What is the geometric structures for N-manifolds of degree 3? In particular, what is the geometric structure corresponding to a split Lie $3$-algebroid?
   \item[$\bullet$] The tangent prolongation of a Lie (Courant) algebroid is a $\VB$-Lie ($\VB$-Courant) algebroid. What are the underlying structures of the tangent prolongation of a split Lie 2 ($\LWX$ 2)-algebroid?

\end{itemize}
We introduce the notions of a $\VB$-Lie 2-algebroid and a $\VB$-$\LWX$ 2-algebroid that   provide answers of above questions.
}

\subsection{Lie $n$-algebroids, Courant algebroids and $\LWX$ 2-algebroids}
An {\bf NQ-manifold} is an N-manifold $\huaM$ together with a degree   $1$ vector field $Q$ satisfying $[Q,Q]=0$.
It is well known that a degree 1 NQ manifold corresponds to a Lie algebroid. Thus, people usually think that
\begin{center}
  An NQ-manifold of degree $n$ corresponds to a Lie $n$-algebroid.
\end{center}
Some work in this direction appeared in
\cite{Voronov:2010halgd}. Strictly speaking, a Lie $n$-algebroid gives
arise to an NQ-manifold only after a degree 1 shift, just as a Lie
algebroid $A$ corresponds to a degree 1 NQ manifold $A[1]$. To make
the shifting manifest, and to present a Lie $n$-algebroid in a way more
used to differential geometers, that is, to use the language of vector
bundles, the authors introduced the notion of a split Lie $n$-algebroid in \cite{sz} to study the integration of a Courant algebroid. The equivalence between the category of split NQ manifolds and the category of split Lie $n$-Lie algebroids was  proved in \cite{BP}. The language of split Lie $n$-algebroids has slowly become a useful tool for differential geometers to study problems related to NQ-manifolds (\cite{thesis,Jotz,Jotz2}). Since Lie 2-algebras are the categorification of Lie algebras (\cite{baez:2algebras}), we will view Lie 2-algebroids as the categorification of Lie algebroids.

To study the double of a Lie bialgebroid (\cite{MackenzieX:1994}), Liu, Weinstein and Xu introduced the notion of a Courant algebroid in \cite{lwx}. See \cite{Roytenbergphdthesis} for an alternative definition. There are many important application of Courant algebroids, e.g. in generalized complex geometry (\cite{BCG,gualtieri,hitchin}), Poisson geometry (\cite{LiM}), moment maps (\cite{BIS}),   Poisson-Lie T-duality (\cite{Sev,SV}) and topological fields theory (\cite{RoyCF}). In \cite{LiuSheng}, the authors introduced the notion of a $\LWX$ 2-algebroid (named after Courant-Liu-Weinstein-Xu), which can be viewed as the categorification of a Courant algebroid. Furthermore, $\LWX$ 2-algebroids are in  one-to-one correspondence with QP-manifolds (symplectic NQ-manifolds) of degree 3, and have applications in the fields theory. See \cite{IU} for more details. 
The underlying algebraic structure of a $\LWX$ 2-algebroid is a Leibniz 2-algebra, or a Lie 3-algebra. There is also a close relationship between $\LWX$ 2-algebroids and the first Pontryagin classes of quadratic Lie 2-algebroids, which are represented by closed 5-forms. More precisely, as the higher analogue of the results given in \cite{Bressler:Pclass,ChenRCA}, it was proved in \cite{sheng} that the first Pontryagin class of a quadratic Lie algebroid  is the obstruction of the existence of a $\LWX$-extension.

\subsection{$\VB$-Lie algebroids and $\VB$-Courant  algebroids}

Double structures in geometry can be traced back to the work of Ehresmann on connection theory, and have been found many applications in Poisson geometry. See \cite{Mac05} for more details.  We use the word ``doublization'' to indicate putting geometric structures on double vector bundles in the sequel. In \cite{Mehta10}, Gracia-Saz and Mehta introduced the notion of a $\VB$-Lie algebroid, which is equivalent to Mackenzie's $\huaL\huaA$-vector bundle (\cite{Mac98}). A $\VB$-Lie algebroid is a Lie algebroid object in the category of vector bundles and one important property is that it is closely related to superconnection (also called representation up to homotopy \cite{abad-crainic:rep-homotopy,AC}) of a Lie algebroid on a 2-term complex of vector bundles. Recently, the relation between $\VB$-algebroid morphism and representation up to homotopy were studied in \cite{DJO}.

In his PhD thesis \cite{Li-Bland}, Li-Bland introduced the notion of a $\VB$-Courant algebroid which is the doublization of a Courant algebroid \cite{lwx}, and established abstract correspondence between NQ-manifolds of degree 2 and  $\VB$-Courant algebroids. Then in \cite{Jotz}, Jotz Lean provided a more concrete description of the equivalence between the category of split Lie 2-algebroids and the category of decomposed  $\VB$-Courant algebroids.

Double structures, such as double principle (vector) bundles (\cite{CLSd,GG,Jotz18,LLL}), double Lie algebroids (\cite{GJMMdouble-ruh,Mac92,Mac98,Mac00,Mac11,Vor12}), double Lie groupoids (\cite{MT}), $\VB$-Lie algebroids (\cite{BC,Mehta10}) and $\VB$-Lie groupoids (\cite{BC,GMehta}) become more and more important recently and are widely studied. In particular, the Lie theory relating $\VB$-Lie algebroids and $\VB$-Lie groupoids, i.e. their relation via differentiation and integration, is established in \cite{BC}.

\subsection{Summary of the results and outline of the paper}
In this paper, we combine the aforementioned higher structures and double structures.  First we introduce the notion of a $\VB$-Lie 2-algebroid, which can be viewed as the categorification of a $\VB$-Lie algebroid, or doublization of a Lie 2-algebroid:
{\footnotesize
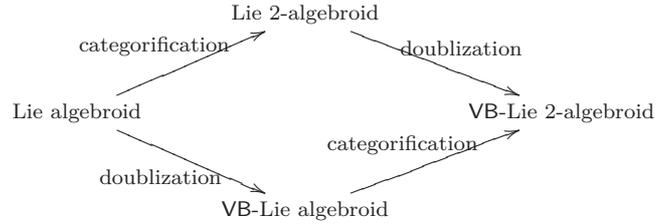
\[
 \xymatrix{ & \mbox{Lie 2-algebroid}\ar[dr]^{\mbox{doublization}}&  \\
\mbox{Lie algebroid}\ar[ur]^{\mbox{categorification}}\ar[dr]_{\mbox{doublization}}&&\VB\mbox{-Lie 2-algebroid.}\\
&\VB\mbox{-Lie algebroid}\ar[ur]^{ \mbox{categorification}}&     }
\]}
We show that the tangent prolongation of a Lie 2-algebroid is a $\VB$-Lie 2-algebroid and the graded fat bundle associated to a $\VB$-Lie 2-algebroid is Lie 2-algebroid. Consequently, the graded jet bundle of a Lie 2-algebroid is also a Lie 2-algebroid. In \cite{Mehta10}, the authors showed that  a $\VB$-Lie algebroid is equivalent to a flat superconnection (representation up to homotopy (\cite{abad-crainic:rep-homotopy})) of a Lie algebroid on a 2-term complex of vector bundles after choosing a splitting. Now for a $\VB$-Lie 2-algebroid, we establish a higher analogous result, namely,  we show that after choosing a splitting, it is equivalent to a flat superconnection of a Lie 2-algebroid on a 3-term complex of vector bundles.

Then we introduce the notion of a $\VB$-$\LWX$ 2-algebroid, which can be viewed as both the doublization of a $\LWX$ 2-algebroid and the categorification of a $\VB$-Courant algebroid. More importantly, we show that after choosing a splitting, there is a one-to-one correspondence between $\VB$-$\LWX$ 2-algebroids and split Lie 3-algebroids (NQ-manifolds of degree 3). The tangent prolongation of a $\LWX$ 2-algebroid is a $\VB$-$\LWX$ 2-algebroid naturally. We go on defining $E$-$\LWX$ 2-algebroid, which can be viewed as the categorification of an $E$-Courant algebroid introduced in \cite{CLS2}. As a higher analogue of the result that associated to a  $\VB$-Courant algebroid, there is an $E$-Courant algebroid \cite{Jotz,LSW}, we show that on the graded fat bundle associated to a $\VB$-$\LWX$ 2-algebroid, there is an $E$-$\LWX$ 2-algebroid structure naturally. Similar to the case of a $\LWX$ 2-algebroid, an $E$-$\LWX$ 2-algebroid also gives rise to a Lie 3-algebra naturally. Thus through the following procedure:
$$
\xymatrix{{}\save[]+<1cm,0cm>*\txt<4pc>{%
Lie 3-algebra }\restore} \longmapsto \xymatrix{{}\save[]+<1cm,0cm>*\txt<6pc>{%
 $\VB$-$\LWX$ 2-algebroid}\restore} \longmapsto \xymatrix{{}\save[]+<1cm,0cm>*\txt<6pc>{%
$E$-$\LWX$ 2-algebroid}\restore} \longmapsto  \xymatrix{{}\save[]+<1cm,0cm>*\txt<4pc>{%
Lie 3-algebra,\restore}}
$$
we can construct a Lie 3-algebra from a Lie 3-algebra. We obtain new interesting examples, including the higher analogue of the string Lie 2-algebra.

The paper is organized as follows. In Section 2, we recall  double vector bundles, $\VB$-Lie algebroids and $\VB$-Courant algebroids. In Section 3, we introduce the notion of a $\VB$-Lie 2-algebroid, and show that both the graded side bundle and the graded fat bundle are Lie 2-algebroids. The tangent prolongation of a Lie 2-algebroid is a $\VB$-Lie 2-algebroid naturally. In Section 4, first we construct a strict Lie 3-algebroid $\End(\huaE)=(\End^{-2}(\huaE),\End^{-1}(\huaE),\dev(\huaE),\frkp,\dM,[\cdot,\cdot]_C)$ from a 3-term complex of vector bundles $\huaE:E_{-2}\stackrel{\pi}{\longrightarrow}E_{-1}\stackrel{\pi}{\longrightarrow}E_{0}$ and then we define a flat superconnection of a Lie 2-algebroid $\huaA=(A_{-1},A_0,\frka,\frkl_1,\frkl_2,\frkl_3)$ on this 3-term complex of vector bundles to be a morphism from $\huaA$ to $\End(\huaE)$. We show that after choosing a splitting, $\VB$-Lie 2-algebroids one-to-one correspond to flat superconnections of a Lie 2-algebroid on a 3-term complex of vector bundles. In Section 5, we introduce the notion of a $\VB$-$\LWX$ 2-algebroid and show that after choosing a splitting, there is a one-to-one correspondence between $\VB$-$\LWX$ 2-algebroids and Lie 3-algebroids. In Section 6, we introduce the notion of an $E$-$\LWX$ 2-algebroid and show that the graded fat bundle associated to a $\VB$-$\LWX$ 2-algebroid  is an $E$-$\LWX$ 2-algebroid naturally. In particular, the graded jet bundle of a $\LWX$ 2-algebroid, which is the graded fat bundle of the tangent prolongation of this $\LWX$ 2-algebroid, is a $T^*M$-$\LWX$ 2-algebroid. We can also obtain a Lie 3-algebra from an $E$-$\LWX$ 2-algebroid. In Section 7, we   construct    a Lie 3-algebra from a given Lie 3-algebra using the theories established in Section 5 and Section 6, and give interesting examples. In particular, we show that associated to a quadratic Lie 2-algebra, we can obtain a Lie 3-algebra, which can be viewed as the higher analogue of the string Lie 2-algebra.

\section{Preliminaries}

See \cite{Mac05} for the precise definition of a double vector bundle. We denote a double vector bundle
\begin{equation*}
\xymatrix{ D \ar@{->} [r]^{\pi_B}\ar@ {->} [d]_{\pi_A}
&B \ar@ {->} [d]^{q_B}&\\
{A} \ar@{->} [r]_{q_A}&M&C \ar@{->} [l]}
\end{equation*}
with core $C$ by $(D;A,B;M)$. We use $D^B$ and $D^A$ to denote  vector bundles $D\longrightarrow B$ and $D\longrightarrow A$ respectively. For a vector bundle $A$, both the tangent bundle $TA$ and the cotangent bundle $T^*A$ are  double vector bundles:
$$
\xymatrix{ TA \ar@{->} [r]^{ }\ar@ {->} [d]_{ }
&TM \ar@ {->} [d]^{ }\\
A \ar@{->} [r]_{ }&M, }
\qquad
\xymatrix{ T^*A \ar@{->} [r]^{ }\ar@ {->} [d]_{ }
&A^* \ar@ {->} [d]^{ }\\
A \ar@{->} [r]_{ }&M. }
$$

A morphism of double vector bundles
    $$ (\varphi;f_A,f_B;f_M):(D;A,B;M)
    \rightarrow(D';A',B';M')$$
consists of maps $\varphi$: $D\rightarrow D'$, $f_A:A\rightarrow
A'$, $f_B:B\rightarrow B'$, $f_M:M\rightarrow M'$, such that each of
$(\varphi,f_B)$, $(\varphi,f_A)$, $(f_A,f_M)$ and $(f_B,f_M)$ is a
morphism of the relevant vector bundles.

The space of sections $\Gamma_B(D)$ of the vector bundle $D^B$ is generated as a $C^\infty(B)$-module by core sections $\Gamma_B^c(D)$ and linear sections $\Gamma_B^l(D)$. See \cite{Mac11} for more details.
For a section $c: M\rightarrow C$, the corresponding {\bf core section} $c^\dag: B\rightarrow D$ is defined as
$$c^\dag(b_m)=\tilde{0}_{b_m}+_A \overline{c(m)},\quad \forall~ m\in M,~ b_m\in B_m,$$
where $\bar{\cdot}$ means the inclusion $C\hookrightarrow D$.
A section $\xi: B\rightarrow D$ is called {\bf linear} if it is a bundle morphism from $B\rightarrow M$ to $D\rightarrow A$ over a section $X\in \Gamma(A)$.
Given $\psi\in \Gamma(B^*\otimes C)$, there is a linear section $\tilde{\psi}: B\rightarrow D$ over the zero section $0^A: M\rightarrow A$ given by
$$\widetilde{\psi}(b_m)=\tilde{0}_{b_m}+_A \overline{\psi(b_m)}.$$

Note that $\Gamma_B^l(D)$ is locally free as a $C^\infty (M)$-module. Therefore, $\Gamma_B^l(D)$ is equal to $\Gamma(\hat{A})$ for some vector bundle $\hat{A}\rightarrow M$. The vector bundle $\hat{A}$ is called the {\bf fat bundle} of the double vector bundle $(D;A,B;M)$. Moreover, we have the following short exact sequence of vector bundles over $M$
\begin{equation}\label{seq:fat}
0 \rightarrow B^*\otimes C \longrightarrow \hat{A}\stackrel{\pr}{\longrightarrow} A \rightarrow 0.
\end{equation}
We will view $ B^*\otimes C$ both as $\Hom(B,C)$ and $\Hom(C^*,B^*)$ depending on what it acts.

\begin{defi}{\rm(\cite{Mehta10})}
  A {\bf $\VB$-Lie algebroid} is a double vector bundle $(D;A,B;M)$ equipped with a Lie algebroid structure $(D^B,a,[\cdot,\cdot]_D)$   such that the anchor $a:D\longrightarrow TB$ is linear, i.e. $a$ is a double vector bundle morphism over $\frka:A\longrightarrow TM$, and the Lie bracket $[\cdot,\cdot]_D$ is linear:
  $$
  [\Gamma_B^l(D),\Gamma_B^l(D)]_D\subset \Gamma_B^l(D),\quad [\Gamma_B^l(D),\Gamma_B^c(D)]_D\subset \Gamma_B^c(D),\quad [ \Gamma_B^c(D), \Gamma_B^c(D)]_D=0.
  $$
\end{defi}
The vector bundle $A\longrightarrow M$ is then also a Lie algebroid, with the anchor $\frka$ and the bracket $[\cdot,\cdot]_A$
defined as follows: if $\xi_1,\xi_2$ are linear over $X_1, X_2\in\Gamma(A)$, then the bracket
$[\xi_1,\xi_2]_D$ is linear over $[X_1, X_2]_A$.

 \begin{defi} {\rm(\cite{Li-Bland})}
A {\bf $\VB$-Courant algebroid} is a metric double vector bundle
$(D;A,B;M)$ such that $(D^ B,S,\Dorfman{\cdot,\cdot},\rho)$ is a Courant algebroid and the following conditions are satisfied:
\begin{itemize}
\item[\rm(i)] The anchor map $\rho:D\rightarrow TB$ is linear, i.e.
$\rho: (D;A,B;M)\longrightarrow (TB;TM,B;M)$
is a morphism of double vector bundles;
\item[\rm(ii)] The Courant bracket is linear. That is
$$\Dorfman{\Gamma_B^l(D),\Gamma_B^l(D)}\subseteq \Gamma_B^l(D),\ \ \ \Dorfman{\Gamma_B^l(D),\Gamma_B^c(D)}\subseteq \Gamma_B^c(D),\quad \Dorfman{\Gamma_B^c(D),\Gamma_B^c(D)}=0.$$
\end{itemize}
\end{defi}

\begin{thm}{\rm(\cite{Li-Bland})}\label{thm:correspondence}
  There  is a one-to-one correspondence between Lie $2$-algebroids and $\VB$-Courant algebroids.
\end{thm}

\emptycomment{
\begin{ex}{\rm
  Let $(A_{-1},A_0,a, l_1,l_2,l_3)$ be a split Lie $2$-algebroid. Then $T^*[2](A_0\oplus A_{-1})[1]$ is a symplectic NQ manifold of degree $2$.  The corresponding $\VB$-Courant algebroid is given by Diagram (i).
  In particular, for a Lie $2$-algebra   $(\g_{-1},\g_0, l_1,l_2,l_3)$, the corresponding $\VB$-Courant algebroid is given by Diagram (ii).

    \begin{equation*}
\xymatrix{ (A_0\oplus A_0^*)\times A_{-1}^* \ar@{->} [r]^{ }\ar@ {->} [d]_{ }
&A^*_{-1} \ar@ {->} [d]^{ }\\
{A_0} \ar@{->} [r]_{ \mbox{{\rm Diagram (i)}}}&M, }\quad  \xymatrix{ (\g_0\oplus \g_0^*)\times \g_{-1}^* \ar@{->} [r]^{ }\ar@ {->} [d]_{ }
&{\g^*_{-1}} \ar@ {->} [d]^{ }\\
\g_0 \ar@{->} [r]_{\mbox{{\rm Diagram (ii)}} }&pt. }
\end{equation*}
}
\end{ex}
}

\emptycomment{
\begin{ex}
  Let $A $ be a Lie algebroid, viewed as a split Lie $2$-algebroid. Then $T^*[2]A[1]$ is a symplectic NQ manifold of degree $2$.  The corresponding $\VB$-Courant algebroid is given by
  \begin{equation*}
  \quad\xymatrix{ A\oplus A^* \ar@{->} [r]^{ }\ar@ {->} [d]_{ }
&{M}\ar@ {->} [d]^{ }\\
A\ar@{->} [r]_{ \mbox{{\rm Diagram I}}}&M, }
\end{equation*}
\end{ex}
}

\section{$\VB$-Lie 2-algebroids}
In this section, we introduce the notion of a  $\VB$-Lie 2-algebroid, which can be viewed as the categorification of a $\VB$-Lie algebroid introduced in \cite{Mehta10}. First we recall the notion of a Lie $n$-algebroid. See \cite{LadaMartin,stasheff:introductionSHLA} for more information of $L_\infty$-algebras.

\begin{defi}{\rm(\cite{sz})}\label{def:n-algd}
  A split Lie $n$-algebroid is a non-positively graded vector bundle $\huaA=A_0\oplus A_{-1}\oplus\cdots\oplus
  A_{-n+1}$ over a manifold $M$ equipped with a bundle map $a:A_0\longrightarrow
  TM$ (called the anchor), and $n+1$ many brackets $l_i:\Gamma(\wedge^i\huaA)\longrightarrow \Gamma(\huaA)$ with
  degree $2-i$ for $1\le i \le n+1$,  such that
\begin{itemize}
  \item[\rm 1.] $\Gamma(\huaA)$ is an $n$-term $L_\infty$-algebra:
$$
\sum_{i+j=k+1}(-1)^{i(j-1)}\sum_{\sigma \in Sh^{-1}_{i, k-i}}\sgn(\sigma)\Ksgn(\sigma)l_j(l_i(X_{\sigma(1)},\cdots,X_{\sigma(i)}),X_{\sigma(i+1)},\cdots,X_{\sigma(k)})=0,
$$
where the summation is taken over all $(i,k-i)$-unshuffles $ Sh^{-1}_{i, k-i}$ with
$i\geq1$ and ``$\Ksgn(\sigma)$'' is the Koszul sign for a permutation $\sigma\in S_k$, i.e.
$$
X_1\wedge  \cdots\wedge X_k=\Ksgn(\sigma)X_{\sigma(1)}\wedge  \cdots\wedge X_{\sigma(k)}.
$$

\item[\rm 2.]  $l_2$ satisfies the Leibniz rule with respect to the anchor $a$:
$$l_2(X^0,fX)=fl_2(X^0,X)+a(X^0)(f)X,\quad\forall~X^0\in\Gamma(A_0),
~f\in C^\infty(M), ~X\in\Gamma(\huaA).$$

\item[\rm 3.] For $i\neq 2$,  $l_i$'s are $ C^\infty(M)$-linear.
\end{itemize}
\end{defi}

 Denote a Lie $n$-algebroid by $(A_{-n+1},\cdots,A_0,a,l_1,\cdots,l_{n+1})$, or simply by $\huaA$. We will only use   a Lie 2-algebroid   $(A_{-1},A_0,a,l_1,l_2,l_3)$ and a Lie 3-algebroid  $(A_{-2},A_{-1},A_0,a,l_1,l_2,l_3,l_4)$. For a Lie $n$-algebroid, we have a generalized Chevalley-Eilenberg complex $(\Gamma(\Symm(\huaA[1])^*),\delta)$.   See \cite{BP,sz} for more details. Then $\huaA[1]$ is an NQ-manifold of degree $n$. A   split Lie $n$-algebroid morphism   $\huaA \to
\huaA'$  can be defined to be a graded vector bundle morphism $f: \Symm(\huaA[1]) \to \Symm(\huaA'[1])$
such that the induced map $f^*: C(\huaA'[1]) \to C(\huaA[1])$ is a morphism  of NQ manifolds. However it is rather complicated to write down a morphism between split Lie $n$-algebroids in terms of vector bundles, anchors and
brackets, please see \cite[Section 4.1]{BP} for such details. We only give explicit formulas of a morphism from a Lie 2-algebroid to a strict Lie $3$-algebroid $(l_3=0,l_4=0)$   and this is what we will use in this paper to define flat superconnections.

\begin{defi}\label{Def2}
Let $\huaA=(A_{-1},A_0,a,l_1,l_{2},l_3)$ be a Lie $2$-algebroid and $\huaA'=(A_{-2}',A_{-1}',A_0',a',l_1',l_{2}')$
a strict Lie $3$-algebroid. A morphism $F$ from $\huaA$ to $\huaA'$ consists of:
\begin{itemize}
\item[$\bullet$] a bundle map $F^{0}:A_{0}\longrightarrow A_{0}'$,
\item[$\bullet$] a bundle map $F^{1}:A_{-1}\longrightarrow A_{-1}',$
\item[$\bullet$] a bundle map $F^{2}_0:\wedge^2 A_{0} \longrightarrow A_{-1}'$,
\item[$\bullet$] a bundle map $F^{2}_1:A_{0}\wedge A_{-1}\longrightarrow A_{-2}',$
\item[$\bullet$] a bundle map $F^{3}: \wedge^3 A_0\longrightarrow A_{-2}',$
\end{itemize}
such that for all $ X^0,Y^0,Z^0,X^0_i\in \Gamma(A_{0})$, $i=1,2,3,4$, $~X^1,Y^1\in \Gamma(A_{-1}),$ we have
\begin{eqnarray*}
a'\circ F^0&=&a,\\
l_1'\circ F_{1}&=&F_{0}\circ l_1,\\
F^{0}l_{2}(X^0,Y^0)-l_{2}'(F^{0}(X^0),F^{0}(Y^0))&=&l_1'F^{2}_0(X^0,Y^0),\\
F^{1}l_{2}(X^0,Y^1)-l_{2}'(F^{0}(X^0),F^{1}(Y^1))&=&F^{2}_0(X^0,l_1(Y^1))-l_1'F^{2}_1 (X^0,Y^1),\\
l_{2}'(F^{1}(X^1),F^{1}(Y^1))&=&F^{2}_1 (l_1(X^1),Y^1)-F^{2}_1 (X^1,l_1(Y^1)),\\
l_{2}'(F^{0}(X^0),F^{2}(Y^0,Z^0))-F^{2}_0 (l_{2}(X^0,Y^0),Z^0)+c.p.&=&F^{1}(l_{3}(X^0,Y^0,Z^0))+l_1'F^{3}(X^0,Y^0,Z^0),
\end{eqnarray*}
\begin{eqnarray*}
&&l_{2}'(F^{0}(X^0),F^2_1(Y^0,Z^1))+l_{2}'(F^{0}(Y^0),F^2_1(Z^1,X^0))
+l_{2}'(F^{1}(Z^1),F^2_0(X^0,Y^0))\\&=&F^2_1(l_{2}(X^0,Y^0),Z^1)+c.p.
+F^3(X^0,Y^0,l_1(Z^1)),
\end{eqnarray*}
 and
\begin{eqnarray*}
&&\sum_{i=1}^4(-1)^{i+1}\Big(F^2_1(X^0_i,l_3(X^0_1,\cdots,\widehat{X^0_i},\cdots X^0_4))+l_2'(F^0(X^0_i),F^3(X^0_1,\cdots,\widehat{X^0_i},\cdots X^0_4))\Big)\\
&&+\sum_{i<j}(-1)^{i+j}\Big(F^3(l_2(X^0_i,X^0_j),X^0_k,X^0_l)+c.p.-\half l_2'(F^2_0(X^0_i,X^0_j),F^2_0(X^0_k,X^0_l))\Big)=0,
\end{eqnarray*}
where $k<l$ and $\{k,l\}\cap \{i,j\}=\emptyset.$
\end{defi}

Let $(A_{-1},A_0,a,l_1,l_2,l_3)$
 be a Lie 2-algebroid.  Then for all $X^0,Y^0\in\Gamma(A_0)$ and $X^1\in\Gamma(A_{-1})$, Lie derivatives
 $L^0_{X^0}:\Gamma(A^*_{-i})\longrightarrow \Gamma(A^*_{-i})$, $i=0,1$,   $L^1_{X^1}:\Gamma(A_{-1}^*)\longrightarrow \Gamma(A^*_0)$ and $L^3_{X^0,Y^0}:\Gamma(A_{-1}^*)\longrightarrow \Gamma(A^*_0)$ are defined by
\begin{equation}\label{eq:L}
\left\{\begin{array}{rcl}
\langle L^0_{X^0}\alpha^0,Y^0\rangle&=&\rho(X^0)\langle Y^0,\alpha^0\rangle-\langle \alpha^0,l_2(X^0,Y^0)\rangle,\\
\langle L^0_{X^0}\alpha^1,Y^1\rangle&=&\rho(X^0)\langle Y^1,\alpha^1\rangle-\langle \alpha^1,l_2(X^0,Y^1)\rangle,\\
\langle L^1_{X^1}\alpha^1,Y^0\rangle&=&-\langle \alpha^1,l_2(X^1,Y^0)\rangle,\\
\langle L^3_{X^0,Y^0}\alpha^1,Z^0\rangle&=&-\langle \alpha^1,l_3(X^0,Y^0,Z^0)\rangle,\end{array}\right.
\end{equation}
for all $  \alpha^0\in\Gamma(A^*_0),~\alpha^1\in\Gamma(A_{-1}^*),~Y^1\in\Gamma(A_{-1}),~Z^0\in\Gamma(A_0)$. If $(\huaA^*[1],\frka,\frkl_1,\frkl_2,\frkl_3)$ is also a split Lie 2-algebroid, we denote by $\huaL^0,\huaL^1,\huaL^3,\delta_*$ the corresponding operations.

A graded double vector bundle consists of a double vector bundle $(D_{-1};A_{-1},B_{-1};M_{-1})$ of degree $-1$ and a double vector bundle  $(D_{0};A_{0},B_{0};M_0)$ of degree $0$:
\begin{equation*}
\xymatrix{ D_{-1} \ar@{->} [r]^{\pi^{B_{-1}}}\ar@ {->} [d]_{\pi^{A_{-1}}}
&B_{-1} \ar@ {->} [d]^{q^{B_{-1}}}&\\
{A_{-1}} \ar@{->} [r]_{q^{A_{-1}}}&M_{-1}&C_{-1}, \ar@{->} [l]}\quad \xymatrix{ D_{0} \ar@{->} [r]^{\pi^{B_{0}}}\ar@ {->} [d]_{\pi^{A_{0}}}
&B_{0} \ar@ {->} [d]^{q^{B_{0}}}&\\
{A_{0}} \ar@{->} [r]_{q^{A_{0}}}&M&C_{0}. \ar@{->} [l]}
\end{equation*}

 We denote a graded double vector bundle by $\left(\begin{array}{ccc}
D_{-1};&A_{-1},B_{-1};&M_{-1}\\
D_{0};&A_{0},B_{0};&M_0
\end{array}\right)$.
Morphisms between graded double vector bundles can be defined in an obvious way.
We will denote by $\huaD$ and $\huaA$ the graded vector bundles $D_0^B\oplus D_{-1}^B$ and $A_0\oplus A_{-1}$ respectively. Now we are ready to introduce the main object in this section.

\begin{defi}
  A $\VB$-Lie $2$-algebroid is a graded double vector bundle $\left(\begin{array}{ccc}
D_{-1};&A_{-1},B;&M\\
D_{0};&A_{0},B;&M
\end{array}\right)$ equipped with a Lie $2$-algebroid structure $(D_{-1}^B,D_0^B,a,l_1,l_2,l_3)$ on $\huaD$ such that
\begin{itemize}
  \item[\rm(i)] The anchor $a:D_0\longrightarrow TB$ is linear, i.e. we have a bundle map $\frka:A_0\longrightarrow TM$ such that $(a;\frka,\id_B;\id_M)$ is a double vector bundle morphism (see Diagram {\rm(i)});
      \item[\rm(ii)]$l_1$ is linear, i.e. we have a bundle map $\frkl_1:A_{-1}\longrightarrow A_0$ such that $(l_1;\frkl_1,\id_B;\id_M)$ is a double vector bundle morphism (see Diagram {\rm(ii)});
  \item[\rm(iii)] $l_2$ is linear, i.e.
  $$\begin{array}{lll}
    l_2(\Gamma^l_B(D_0),\Gamma^l_B(D_0))\subset \Gamma^l_B(D_0),& l_2(\Gamma^l_B(D_0),\Gamma^c_B(D_0))\subset \Gamma^c_B(D_0),& l_2(\Gamma^c_B(D_0),\Gamma^c_B(D_0))=0,\\
        l_2(\Gamma^l_B(D_0),\Gamma^l_B(D_{-1}))\subset \Gamma^l_B(D_{-1}),& l_2(\Gamma^l_B(D_0),\Gamma^c_B(D_{-1}))\subset \Gamma^c_B(D_{-1}),&\\
      l_2(\Gamma^c_B(D_0),\Gamma^l_B(D_{-1}))\subset \Gamma^c_B(D_{-1}),&
       l_2(\Gamma^c_B(D_0),\Gamma^c_B(D_{-1}))=0;&
  \end{array}
  $$

  \item[\rm(iv)] $l_3$ is linear, i.e.
  \begin{eqnarray*}
 && l_3(\Gamma^l_B(D_0),\Gamma^l_B(D_0),\Gamma^l_B(D_0))\subset \Gamma^l_B(D_{-1}),\quad l_3(\Gamma^l_B(D_0),\Gamma^l_B(D_0),\Gamma^c_B(D_0))\subset \Gamma^c_B(D_{-1}),\\
  &&l_3(\Gamma^c_B(D_0),\Gamma^c_B(D_0),\cdot)=0.
\end{eqnarray*}
\end{itemize}
  \begin{equation*}
\xymatrix@!0{  D_0 \ar@{->} [rr] \ar@ {->} [dd]_{a} \ar@{->} [rd]
&&  B  \ar@{=}[dd] \ar@{->} [rd]&&&\\
&A_0\ar@{->} [rr]\ar@{->} [dd]_{\frka}&&M\ar@{=} [dd]&&C_0\ar@{->}[ll]\\
TB \ar@{->} [rd]&\ar@{->} [r]&B \ar@{->} [rd]& &&\\
&TM\ar@{->} [rr]&&M&& B,\ar@{->}[ll]\\
&&\mbox{\rm Diagram (i)}}  \qquad
 \xymatrix@!0{  D_{-1} \ar@{->} [rr] \ar@ {->} [dd]_{l_1} \ar@{->} [rd]
&&  B  \ar@{=}[dd] \ar@{->} [rd]&&&\\
&A_{-1}\ar@{->} [rr]\ar@{->} [dd]_{\frkl_1}&&M\ar@{=} [dd]&&C_{-1}\ar@{->}[ll]\\
D_0\ar@{->} [rd]& \ar@{->} [r] &B \ar@{->} [rd]& &&\\
&A_0\ar@{->} [rr]&&M&& C_0.\ar@{->}[ll]\\
&&\mbox{\rm Diagram (ii)}}
\end{equation*}
\end{defi}

Since Lie 2-algebroids are the categorification of Lie algebroids, $\VB$-Lie 2-algebroids can be viewed as the categorification of $\VB$-Lie algebroids.

Recall that  if $(D;A,B;M)$ is a $\VB$-Lie algebroid, then $A$ is a Lie algebroid. The following result is its higher analogue.
\begin{thm}\label{thm:sideLie2}
Let $\left(\begin{array}{ccc}
D_{-1};&A_{-1},B;&M\\
D_{0};&A_{0},B;&M
\end{array}\right)$ be a $\VB$-Lie $2$-algebroid. Then $(A_{-1},A_0,\frka,\frkl_1,\frkl_2,\frkl_3)$ is a Lie $2$-algebroid, where $\frkl_2$ is defined by the property that if $\xi^0_1,~\xi^0_2,~\xi^0 \in\Gamma^l_B(D_0)$ are linear sections over $X^0_1,~X^0_2,~X^0\in\Gamma(A_0)$, and $\xi^1\in\Gamma^l_B(D_{-1})$ is a linear section over $X^1\in\Gamma(A_{-1})$, then
$l_2(\xi^0_1,\xi^0_2)\in\Gamma^l_B(D_0)$ is a linear section over $\frkl_2(X^0_1,X^0_2)\in\Gamma(A_0)$ and $l_2(\xi^0 ,\xi^1)\in \Gamma^l_B(D_{-1})$ is a linear section over $\frkl_2(X^0 ,X^1)\in\Gamma(A_{-1})$.
Similarly, $\frkl_3$ is defined by the property that if $\xi^0_1,~\xi^0_2,~\xi^0_3\in\Gamma^l_B(D_0)$ are linear sections over $X^0_1,~X^0_2,~X^0_3\in\Gamma(A_0)$, then $l_3(\xi^0_1,\xi^0_2,\xi^0_3)\in\Gamma^l_B(D_{-1})$ is a linear section over $\frkl_3(X^0_1,X^0_2,X^0_3)\in\Gamma(A_{-1})$.
\end{thm}
\pf Since $l_2$ is linear, for any $\xi^i\in\Gamma^l_B(D_{-i})$ satisfying $\pi^{A_{-i}}(\xi^i)=0$, we have $$\pi^{A_{-(i+j)}}(l_2(\xi^i,\eta^j))=0,\quad\forall~ \eta^j\in\Gamma^l_B(D_{-j}).$$ This implies that $\frkl_2$ is well-defined. Similarly, $\frkl_3$ is also well-defined.

By the fact that $l_1:D_{-1}\longrightarrow D_0$ is a double vector bundle morphism over $\frkl_1:A_{-1}\longrightarrow A_0$, we can deduce that $(\Gamma(A_{-1}),\Gamma(A_0),\frkl_1,\frkl_2,\frkl_3)$ is a Lie 2-algebra. We only give a proof of the property \begin{equation}\label{eq:con1Lie2}
\frkl_1(\frkl_2(X_0,X_1))=\frkl_2(X_0,\frkl_1 (X_1)),\quad \forall X^0\in\Gamma(A_0), ~X^1\in\Gamma(A_{-1}).
 \end{equation}The other conditions in the definition of a Lie 2-algebra can be proved similarly. In fact, let $\xi^0\in\Gamma^l_B(D_0),~\xi^1\in\Gamma^l_B(D_{-1})$ be linear sections over $X^0,X^1$ respectively, then by the equality $l_1(l_2(\xi^0,\xi^1))=l_2(\xi^0,l_1(\xi^1))$, we have
$$
\pi^{A_{0}}l_1(l_2(\xi^0,\xi^1))=\pi^{A_{0}}l_2(\xi^0,l_1(\xi^1)).
$$
Since $l_1:D_{-1}\longrightarrow D_0$ is a double vector bundle morphism over $\frkl_1:A_{-1}\longrightarrow A_0$, the left hand side is equal to
$$
\pi^{A_{0}}l_1(l_2(\xi^0,\xi^1))=\frkl_1\pi^{A_{-1}}l_2(\xi^0, \xi^1)=\frkl_1\frkl_2(X^0,X^1),
$$
and the right hand side is equal to
$$
\pi^{A_{0}}l_2(\xi^0,l_1(\xi^1))=\frkl_2(\pi^{A_{0}}(\xi^0),\pi^{A_{0}}(l_1(\xi^1)))=\frkl_2(X_0,\frkl_1(X^1)).
$$
Thus, we deduce that \eqref{eq:con1Lie2} holds.

Finally, for all $X^0\in\Gamma(A_0)$, $Y^i\in\Gamma(A_{-i})$ and $f\in\CWM$, let $\xi^0\in\Gamma^l_B(D_0)$ and $\eta^i\in\Gamma^l_B(D_{-i}),~i=0,1$ be linear sections over $X^0$ and $Y^i$. Then $q_B^*(f)\eta^i$ is a linear section over $fY^i$. By the fact that $a$ is a double vector bundle morphism over $\frka$, we have
\begin{eqnarray*}
  \frkl_2(X^0,fY^i)&=&\pi^{A_{-i}}l_2(\xi^0,q_B^*(f)\eta^i)=\pi^{A_{-i}}\big(q_B^*(f)l_2(\xi^0,\eta^i)+a(\xi^0)(q_B^*(f))\eta^i\big)\\
  &=&f\frkl_2(X^0,Y^i)+\frka(X^0)(f)Y^i.
\end{eqnarray*}
Therefore,  $(A_{-1},A_0,\frka,\frkl_1,\frkl_2,\frkl_3)$ is a Lie $2$-algebroid.\qed

\begin{rmk}
  By the above theorem, we can view a $\VB$-Lie $2$-algebroid as a Lie $2$-algebroid object in the category of double vector bundles.
\end{rmk}

Consider the associated graded fat bundle $\hat{A}_{-1}\oplus \hat{A}_0$, obviously we have
\begin{pro}\label{pro:fatLie2}
Let $\left(\begin{array}{ccc}
D_{-1};&A_{-1},B;&M\\
D_{0};&A_{0},B;&M
\end{array}\right)$ be a $\VB$-Lie $2$-algebroid. Then $(\hat{A}_{-1}, \hat{A}_0,\hat{a},\hat{l}_1,\hat{l}_2,\hat{l}_3)$ is a Lie $2$-algebroid, where $\hat{a}=\frka\circ \pr $ and $\hat{l}_1,~\hat{l}_2,~\hat{l}_3$ are the restriction of $l_1,~l_2,~l_3$ on linear sections respectively.

Consequently, we have the following exact sequences of Lie $2$-algebroids:
\begin{equation}\label{seq:Lie2}
\CD
  0 @>   >>  B^*\otimes C_{-1} @>  >> \hat{A}_{-1}  @>\pr   >> A_{-1} @>   >> 0 \\
  @.     @V   VV @V \hat{l}_1 VV @V\frkl_1 VV @.  \\
  0 @>   >> B^*\otimes C_{0} @>  >> \hat{A}_0 @>\pr>> A_0@>   >>0
\endCD
\end{equation}
\end{pro}

 It is helpful to give the Lie 2-algebroid structure on $B^*\otimes C_{-1}\oplus B^*\otimes C_0$. Since $l_1$ is linear, it induces a bundle map   $l_1^C:C_{-1}\longrightarrow C_0$. The restriction of $\hat{l}_1$ on $B^*\otimes C_{-1}$ is given by
\begin{equation}\label{eq:Lie2ab1}
  \hat{l}_1(\phi^1)=l_1^C\circ \phi^1,\quad\forall \phi^1\in\Gamma(B^*\otimes C_{-1})=\Gamma(\Hom(B,C_{-1})).
\end{equation}

 Since the anchor $a:D_0\longrightarrow TB$ is a double vector bundle morphism, it induces a bundle map $\varrho:C_0\longrightarrow B$ via
 \begin{equation}
   \langle \varrho(c^0),\xi\rangle =-a(c^0)(\xi),\quad\forall c^0\in\Gamma(C_0),~ \xi\in \Gamma(B^*).
 \end{equation}
Then by the Leibniz rule, we   deduce that the restriction of $\hat{l}_2$ on $\Gamma(B^*\otimes C_{-1}\oplus B^*\otimes C_0)$ is given by
\begin{eqnarray}
  \label{eq:Lie2ab2}\hat{l}_2(\phi^0,\psi^0)&=&\phi^0\circ \varrho\circ\psi^0-\psi^0\circ \varrho\circ\phi^0,\\
  \label{eq:Lie2ab3} \hat{l}_2(\phi^0,\psi^1)&=&- \hat{l}_2(\psi^1,\phi^0)=-\psi^1\circ \varrho\circ\phi^0,
\end{eqnarray}
for all $\phi^0,\psi^0\in\Gamma(B^*\otimes C_{0})=\Gamma(\Hom(B,C_{0}))$ and $\psi^1\in \Gamma(B^*\otimes C_{-1})=\Gamma(\Hom(B,C_{-1}))$. Since $l_3$ is linear, the restriction of $l_3$ on $B^*\otimes C_{-1}\oplus B^*\otimes C_0$ vanishes. Obviously, the anchor is trivial. Thus, the Lie 2-algebroid structure on $B^*\otimes C_{-1}\oplus B^*\otimes C_0$ is exactly given by \eqref{eq:Lie2ab1}, \eqref{eq:Lie2ab2} and \eqref{eq:Lie2ab3}. Therefore, $B^*\otimes C_{-1}\oplus B^*\otimes C_0$ is a graded bundle of strict Lie 2-algebras.

\vspace{3mm}

An important example of $\VB$-Lie algebroids is the tangent prolongation of a Lie algebroid. Now we explore the tangent prolongation of a Lie 2-algebroid. Recall that for a Lie algebroid $A\stackrel{ }{\longrightarrow}M$, $TA$ is a Lie algebroid over $TM$.  A section $\sigma:M\longrightarrow A$   gives rise to a linear section $\sigma_T\triangleq d\sigma:TM\longrightarrow TA$ and a core section  $\sigma_C:TM\longrightarrow TA$ by contraction. Any section of $TA$ over $TM$ is generated by such sections. A function $f\in \CWM$  induces two types of functions on $TM$ by
$$
f_C=q^*f,\quad f_T=df,
$$
where $q:TM\longrightarrow M$ is the projection.  We have the following relations about the module structure:
\begin{equation}
  (f\sigma)_C=f_C\sigma_C,\quad (f\sigma)_T=f_T\sigma_C+f_C\sigma_T.
\end{equation}
In particular, for $A=TM$, we have
\begin{equation}
  X_T(f_T)=X(f)_T,\quad X_T(f_C)=X(f)_C,\quad X_C(f_T)=X(f)_C,\quad X_C(f_C)=0,\quad \forall X\in\frkX(M).
\end{equation}
See \cite[Example 2.5.4]{Li-Bland} and \cite{Mac05} for more details.

Now for Lie 2-algebroids, we have

\begin{pro}\label{pro:tangentpLie2}
Let $\huaA=( A_{-1},A_0,\frka,\frkl_1,\frkl_2,\frkl_3)$ be a Lie $2$-algebroid. Then $(TA_{-1},TA_0,a,l_1,l_2,l_3)$ is a Lie $2$-algebroid over $TM$, where $a:TA_0\longrightarrow TTM$ is given by
\begin{equation}
  a(\sigma^0_T)=\frka(\sigma^0)_T,\quad  a(\sigma^0_C)=\frka(\sigma^0)_C,
\end{equation}
$l_1:\Gamma_{TM}(TA_{-1})\longrightarrow \Gamma_{TM}(TA_0)$ is given by
\begin{equation}
  l_1(\sigma^1_T)=\frkl_1(\sigma^1)_T,\quad  l_1(\sigma^1_C)=\frkl_1(\sigma^1)_C,
\end{equation}
 $l_2:\Gamma_{TM}(TA_{-i})\times \Gamma_{TM}(TA_{-j})\longrightarrow \Gamma_{TM}(TA_{-(i+j)})$ is given by
\begin{eqnarray*}
 && l_2(\sigma^0_T,\tau^0_T)=\frkl_2(\sigma^0,\tau^0)_T,~  l_2(\sigma^0_T,\tau^0_C)=\frkl_2(\sigma^0,\tau^0)_C,~l_2(\sigma^0_C,\tau^0_C)=0,\\
  && l_2(\sigma^0_T,\tau^1_T)=\frkl_2(\sigma^0,\tau^1)_T,~  l_2(\sigma^0_T,\tau^1_C)=\frkl_2(\sigma^0,\tau^1)_C,~ l_2(\sigma^0_C,\tau^1_T)=\frkl_2(\sigma^0,\tau^1)_C,~ l_2(\sigma^0_C,\tau^1_C)=0,
\end{eqnarray*}
and $l_3:\wedge^3\Gamma_{TM}(TA_{0})\longrightarrow \Gamma_{TM}(TA_{-1})$ is given by
\begin{equation}
  l_3(\sigma^0_T,\tau^0_T,\varsigma^0_T)=\frkl_3(\sigma^0,\tau^0,\varsigma^0)_T,\quad l_3(\sigma^0_T,\tau^0_T,\varsigma^0_C)=\frkl_3(\sigma^0,\tau^0,\varsigma^0)_C,\quad l_3(\sigma^0_T,\tau^0_C,\varsigma^0_C)=0,
\end{equation}
 for all $\sigma^0,\tau^0,\varsigma^0\in\Gamma(A_0)$ and $\sigma^1,\tau^1\in\Gamma(A_{-1}).$

 Moreover,  we have  the following $\VB$-Lie $2$-algebroid:
\begin{equation*}
\xymatrix@!0{  TA_{-1} \ar@{->} [rr] \ar@ {->} [dd] \ar@{->} [rd]
&&  TM  \ar@ {->} [dd] \ar@{->} [rd]&&&\\
&A_{-1}\ar@{->} [rr]\ar@{->} [dd]&&M\ar@{->} [dd]& &A_{-1}\ar@{->}[ll]\\
TA_0\ar'[r][rr] \ar@{->} [rd]&&TM \ar@{->} [rd]& &&\\
&A_0\ar@{->} [rr]&&M& &A_0.\ar@{->}[ll]}
\end{equation*}
\end{pro}
\pf By the fact that $\huaA=( A_{-1},A_0,\frka,\frkl_1,\frkl_2,\frkl_3)$ is a Lie $2$-algebroid, it is straightforward to deduce that $(TA_{-1},TA_0,a,l_1,l_2,l_3)$ is a Lie $2$-algebroid over $TM$. Moreover, $a,~l_1,~l_2,~l_3$ are all linear, which implies that we have a $\VB$-Lie 2-algebroid. \qed\vspace{3mm}

  The associated fat bundles of double vector bundles $(TA_{-1};A_{-1},TM;M)$ and $(TA_{0};A_{0},TM;M)$ are the jet bundles $\jet A_{-1}$ and $\jet A_0$ respectively. By Proposition \ref{pro:tangentpLie2} and Proposition \ref{pro:fatLie2}, we obtain the following result, which is the higher analogue of the fact that the jet bundle of a Lie algebroid is a Lie algebroid.

\begin{cor}
Let $( A_{-1},A_0,\frka,\frkl_1,\frkl_2,\frkl_3)$ be a Lie $2$-algebroid. Then $(\jet A_{-1},\jet A_0,\hat{a},\hat{l}_1,\hat{l}_2,\hat{l}_3)$ is a Lie $2$-algebroid, where $\hat{a},~\hat{l}_1,~\hat{l}_2,~\hat{l}_3$ is given by
\begin{eqnarray*}
  \hat{a}(\sigma^0_T)&=&\frka(\sigma^0),\\
  \hat{l}_2(\sigma^0_T,\tau^0_T)&=&\frkl_2(\sigma^0,\tau^0)_T,\\
  \hat{l}_2(\sigma^0_T,\tau^1_T)&=&\frkl_2(\sigma^0,\tau^1)_T,\\
  \hat{l}_3(\sigma^0_T,\tau^0_T,\zeta^0_T)&=&\frkl_2(\sigma^0,\tau^0,\zeta^0)_T,
\end{eqnarray*}
for all $\sigma^0,~\tau^0,~\zeta^0\in\Gamma(A_0)$ and $\tau^1\in\Gamma(A_{-1})$.
\end{cor}

\section{Superconnections of a Lie $2$-algebroid on a $3$-term complex of vector bundles}

In the section, we introduce the notion of a superconnection of a Lie $2$-algebroid on a $3$-term complex of vector bundles, which generalizes the notion of a superconnection of a Lie  algebroid on a $2$-term complex of vector bundles studied in \cite{Mehta10}. We show that a $\VB$-Lie $2$-algebroid structure on a split graded double vector bundle is equivalent to a  flat superconnection of a Lie $2$-algebroid on a $3$-term complex of vector bundles.

Denote a 3-term complex of vector bundles $E_{-2}\stackrel{\pi}{\longrightarrow}E_{-1}\stackrel{\pi}{\longrightarrow}E_{0}$ by $\huaE$. Then the covariant differential operator bundle $\dev(\huaE)$ fits the following exact sequence:
\begin{equation}
  0\stackrel{}{\longrightarrow}\End(E_0)\oplus \End(E_{-1})\oplus \End(E_{-2})\stackrel{}{\longrightarrow} \dev(\huaE)\stackrel{}{\longrightarrow}TM\stackrel{}{\longrightarrow}0.
\end{equation}
Sections of $\dev(\huaE)$ are of the form $\frkd=(\frkd_0,\frkd_1,\frkd_2)$, where $\frkd_i:\Gamma(E_{-i})\longrightarrow\Gamma(E_{-i})$ are $\mathbb R$-linear maps such that there exists $X\in\frkX(M)$ satisfying
$$
\frkd_i(fe^i)=f\frkd_i(e^i)+X(f)e^i,\quad \forall f\in\CWM,~ e^i\in\Gamma(E_{-i}).
$$
Equivalently, $\dev(\huaE)=\dev(E_0)\times_{TM}\dev(E_{-1})\times_{TM}\dev(E_{-2})$.
Define $\frkp:\dev(\huaE)\longrightarrow TM$ by
\begin{equation}
  \frkp(\frkd_0,\frkd_1,\frkd_2)=X.
\end{equation}
Denote by $\End^{-1}(\huaE)=\Hom(E_0,E_{-1})\oplus \Hom(E_{-1},E_{-2})$ and $\End^{-2}(\huaE)=\Hom(E_0,E_{-2})$. Define $\dM:\End^{-2}(\huaE)\longrightarrow\End^{-1}(\huaE)$ by
\begin{equation}
  \dM(\theta^2)=\pi\circ \theta^2-\theta^2\circ \pi,\quad \forall \theta^2\in\Gamma(\Hom(E_0,E_{-2})),
\end{equation}
and define $\dM:\End^{-1}(\huaE)\longrightarrow\dev(\huaE)$ by
\begin{equation}
  \dM(\theta^1)=\pi\circ \theta^1+\theta^1\circ \pi,\quad \forall \theta^1\in\Gamma(\Hom(E_0,E_{-1})\oplus \Hom(E_{-1},E_{-2})).
\end{equation}
Then we define a degree $0$ graded symmetric bracket operation $[\cdot,\cdot]_C$ on the section space of the graded bundle $\End^{-2}(\huaE)\oplus \End^{-1}(\huaE)\oplus \dev(\huaE)$ by
\begin{eqnarray}
~[\frkd,\frkt]_C&=&\frkd\circ\frkt-\frkt\circ \frkd,\quad \forall \frkd, \frkt\in\Gamma(\dev(\huaE)),\\
  ~[\frkd,\theta^i]_C&=&\frkd\circ \theta^i-\theta^i\circ\frkd,\quad\forall \frkd\in\Gamma(\dev(\huaE)),~\theta^i\in\Gamma(\End^{-i}(\huaE)),\\
  ~[\theta^1,\vartheta^1]_C&=&\theta^1\circ \vartheta^1+\vartheta^1\circ\theta^1,\quad\forall \theta^1,\vartheta^1\in\Gamma(\End^{-1}(\huaE)).
\end{eqnarray}

Denote by $\dev_\pi(\huaE)\subset \dev(\huaE)$ the subbundle of $\dev(\huaE)$ whose section $\frkd\in\Gamma(\dev_\pi(\huaE))$ satisfying $\pi\circ\frkd=\frkd\circ\pi$, or in term of components,
$$
\frkd_0\circ\pi=\pi\circ\frkd_1,\quad \frkd_1\circ\pi=\pi\circ\frkd_2.
$$
It is obvious that $\Gamma(\dev_\pi(\huaE))$ is closed under the bracket operation $[\cdot,\cdot]_C$ and $\dM(\End^{-1}(\huaE))\subset \dev_\pi(\huaE)$.
Then it is straightforward to verify that
\begin{thm}
 Let $E_{-2}\stackrel{\pi}{\longrightarrow}E_{-1}\stackrel{\pi}{\longrightarrow}E_{0}$ be a $3$-term complex of vector bundles  over $M$. Then     $(\End^{-2}(\huaE),\End^{-1}(\huaE),\dev_\pi(\huaE),\frkp,\dM,[\cdot,\cdot]_C)$ is a strict Lie $3$-algebroid.
\end{thm}

With above preparations, we give the definition of a superconnection of a Lie 2-algebroid on a $3$-term complex of vector bundles as follows.

\begin{defi}
   A {\bf superconnection} of a Lie $2$-algebroid $(A_{-1},A_0,\frka,\frkl_1,\frkl_2,\frkl_3)$ on   a $3$-term complex of vector bundles $E_{-2}\stackrel{\pi}{\longrightarrow}E_{-1}\stackrel{\pi}{\longrightarrow}E_{0}$ consists of:
   \begin{itemize}
     \item[$\bullet$] a bundle morphism $F^0:A_0\longrightarrow\dev_\pi(\huaE)$,
     \item[$\bullet$] a bundle morphism $F^1:A_{-1}\longrightarrow\End^{-1}(\huaE)$,
     \item[$\bullet$] a bundle morphism $F^2_0:\wedge^2A_{0}\longrightarrow\End^{-1}(\huaE)$,
     \item[$\bullet$] a bundle morphism $F^2_1: A_{0}\wedge A_{-1}\longrightarrow\End^{-2}(\huaE)$,
       \item[$\bullet$] a bundle morphism $F^3: \wedge^3 A_{0} \longrightarrow\End^{-2}(\huaE)$.
   \end{itemize}
   A superconnection is called {\bf flat} if $(F^0,F^1,F^2_0,F^2_1,F^3)$ is a Lie $n$-algebroid morphism from the Lie $2$ -algebroid $(A_{-1},A_0,\frka,\frkl_1,\frkl_2,\frkl_3)$ to the strict Lie $3$-algebroid $(\End^{-2}(\huaE),\End^{-1}(\huaE),\dev_\pi(\huaE),\frkp,\dM,[\cdot,\cdot]_C).$
\end{defi}

\begin{rmk}
  If the Lie $2$-algebroid reduces to a Lie algebroid $A$ and the $3$-term complex reduces to a $2$-term complex $E_{-1}\stackrel{\pi}{\longrightarrow} E_0$, a superconnection will only consists of
  \begin{itemize}
     \item[$\bullet$] a bundle morphism $F^0=(F^0_0,F^0_1):A\longrightarrow\dev_\pi(\huaE)$,
         \item[$\bullet$] a bundle morphism $F^2_0:\wedge^2A_{0}\longrightarrow\Hom(E_0,E_{-1})$.
        \end{itemize}
        Thus, we recover the notion of a superconnection (also called representation up to homotopy if it is flat) of a Lie algebroid on a $2$-term complex of vector bundles. See \cite{abad-crainic:rep-homotopy,Mehta10}  for more details.
\end{rmk}

Now we come back to $\VB$-Lie $2$-algebroids. Let $(D_{-1}^B,D_0^B,a,l_1,l_2,l_3)$  be a $\VB$-Lie $2$-algebroid structure on the graded double vector bundle $\left(\begin{array}{ccc}
D_{-1};&A_{-1},B;&M\\
D_{0};&A_{0},B;&M
\end{array}\right)$. Recall from Theorem \ref{thm:sideLie2} and Proposition \ref{pro:fatLie2} that both $(A_{-1},A_0,\frka,\frkl_1,\frkl_2,\frkl_3)$ and $(\hat{A}_{-1}, \hat{A}_0,\hat{a},\hat{l}_1,\hat{l}_2,\hat{l}_3)$ are Lie $2$-algebroids.

Choose a horizontal lift $s=(s_0,s_1):A_0\oplus A_{-1}\longrightarrow \hat{A}_0\oplus \hat{A}_{-1}$ of the short exact sequence of Lie 2-algebroids \eqref{seq:Lie2}. Define $\nabla^B:A_0\longrightarrow \dev(B)$ by
\begin{equation}
 \langle \nabla^B_{X^0}b,\xi\rangle=\frka(X^0)\langle\xi,b\rangle-\langle b,\hat{a}(s_0(X^0))(\xi)\rangle,\quad\forall X^0\in\Gamma(A_0),~b\in\Gamma(B),~\xi\in\Gamma(B^*).
\end{equation}
Since for all $\phi^0\in\Gamma(B^*\otimes C_0)$, we have $\hat{a}(\phi^0)=0$, it follows that $\nabla^B$ is well-defined.

We define $\nabla^0:A_0\longrightarrow \dev(C_0)$ and $\nabla^1:A_0\longrightarrow \dev(C_{-1})$ by
\begin{eqnarray}
  \nabla^0_{X^0}c^0=l_2(s_0(X^0),c^0),\quad   \nabla^1_{X^0}c^1=l_2(s_0(X^0),c^1),\quad \forall X^0\in\Gamma(A_0),~c^0\in\Gamma(C_0),~c^1\in\Gamma(C_{-1}).
\end{eqnarray}

Define $\Upsilon^1:A_{-1}\longrightarrow \Hom(B,C_0)$ and $\Upsilon^2:A_{-1}\longrightarrow \Hom(C_0,C_{-1})$ by
\begin{equation}
 \Upsilon^1_{X^1}=s_0(\frkl_1(X^1))-\hat{l}_1(s_1(X^1)),\quad \Upsilon^2_{X^1}c^0=l_2(s_1(X^1),c^0),\quad \forall X^1\in\Gamma(A_{-1}),~c^0\in\Gamma(C_0).
\end{equation}
Since $l_2$ is linear, $\nabla^0$, $\nabla^1$ and $\Upsilon$ are well-defined.

Define $R^0:\wedge^2\Gamma(A_0)\longrightarrow \Gamma(\Hom(B,C_0))$, $\Lambda:\wedge^2\Gamma(A_0)\longrightarrow \Gamma(\Hom(C_0,C_{-1}))$ and $R^1:\Gamma(A_0)\wedge\Gamma(A_{-1})\longrightarrow\Gamma(\Hom(B,C_{-1}))$ by
\begin{eqnarray}
  R^0(X^0,Y^0)&=&s_0\frkl_2(X^0,Y^0)-\hat{l}_2(s_0(X^0),s_0(Y^0)),\\
  \Lambda(X^0,Y^0)(c^0)&=&- {l}_3(s_0(X^0),s_0(Y^0),c^0),\\
  R^1(X^0,Y^1)&=&s_1\frkl_2(X^0,Y^1)-\hat{l}_2(s_0(X^0),s_1(Y^1)),
\end{eqnarray}
for all $X^0,~Y^0\in\Gamma(A_0)$ and $Y^1\in \Gamma(A_{-1})$

Finally, define $\Xi:\wedge^3\Gamma(A_0)\longrightarrow \Hom(B,C_{-1})$ by
\begin{equation}
  \Xi(X^0,Y^0,Z^0))=s_1\frkl_3(X^0,Y^0,Z^0)-\hat{l}_3(s_0(X^0),s_0(Y^0),s_0(Z^0)),\quad\forall X^0,Y^0,Z^0\in\Gamma(A_0).
\end{equation}

By the equality $l_1l_2(s_0(X^0),c^1)=l_2(s_0(X^0),l_1^C(c^1))$, we obtain
\begin{equation}\label{eq:comm1}
  l_1^C\circ \nabla_{X^0}^1=\nabla_{X^0}^0\circ l_1^C.
\end{equation}
By the fact that $a:D_0\longrightarrow TB$ preserves the bracket operation, we obtain
\begin{eqnarray*}
 \langle \nabla^B_{X^0}\varrho(c^0),\xi\rangle &=&\frka(X^0)\langle \varrho(c^0),\xi\rangle-\langle\varrho(c^0),a(s_0(X^0))(\xi)\rangle\\
 &=&-[a(s_0(X^0)),a(c^0)]_{TB}(\xi)=-a\big(l_2(s_0(X^0),c^0)\big)(\xi)=\langle\varrho\nabla^0_{X^0} c^0 ,\xi\rangle,
\end{eqnarray*}
which implies that \begin{equation}\label{eq:comm2}
\nabla^B_{X^0}\circ\varrho=\varrho\circ \nabla^0_{X^0}.
\end{equation}
By \eqref{eq:comm1} and \eqref{eq:comm2}, we deduce that $(\nabla^B_{X^0},\nabla^0_{X^0},\nabla^1_{X^0})\in\dev (\huaE)$, where $\huaE$ is the 3-term complex of vector bundles $C_{-1}\stackrel{l_1^C}{\longrightarrow}C_{0}\stackrel{\varrho}{\longrightarrow}B$ . Then we obtain a superconnection $(F^0,F^1,F^2_0,F^2_1,F^3)$ of the Lie 2-algebroid $(A_{-1},A_0,\frka,\frkl_1,\frkl_2,\frkl_3)$ on the $3$-term complex of vector bundles $C_{-1}\stackrel{l_1^C}{\longrightarrow}C_{0}\stackrel{\varrho}{\longrightarrow}B$, where
$$
F^0=(\nabla^B,\nabla^0,\nabla^1),\quad F^1=(\Upsilon^1,\Upsilon^2),\quad F^2_0=(R^0,\Lambda),\quad F^2_1=R^1,\quad F^3=\Xi.
$$

\begin{thm}
 After choosing a horizontal lift $s=(s_0,s_1):A_0\oplus A_{-1}\longrightarrow \hat{A}_0\oplus \hat{A}_{-1}$, there is a one-to-one correspondence between $\VB$-Lie $2$-algebroids $\left(\begin{array}{ccc}
D_{-1};&A_{-1},B;&M\\
D_{0};&A_{0},B;&M
\end{array}\right)$ and flat superconnections $(F^0,F^1,F^2_0,F^2_1,F^3)$ of the Lie $2$-algebroid  $(A_{-1},A_0,\frka,\frkl_1,\frkl_2,\frkl_3)$ on the $3$-term complex of vector bundles $C_{-1}\stackrel{l_1^C}{\longrightarrow}C_{0}\stackrel{\varrho}{\longrightarrow}B$.
\end{thm}
\pf First it is obvious that
\begin{equation}\label{eq:anccom}
  \frkp\circ F^0=\frka.
\end{equation}

Using equalities $\frka\circ \frkl_1=0$ and $a\circ l_1=0$, we have
\begin{eqnarray*}
  \langle\nabla^B_{\frkl_1X^1}b,\xi\rangle&=&\frka(\frkl_1(X^1))\langle b,\xi\rangle-\langle b,a(s_0(\frkl_1(X^1)))(\xi)\rangle=-\langle b,a(\Upsilon^1_{X^1})(\xi)\rangle,
\end{eqnarray*}
which implies that
\begin{equation}\label{eq:diff1}
\nabla^B_{\frkl_1X^1}=\varrho\circ \Upsilon^1_{X^1}.
\end{equation}
For $\nabla^0$, we can obtain
\begin{eqnarray}\label{eq:diff2}
  \nabla^0_{\frkl_1(X^1)}=l_2(s_0\frkl_1(X^1),\cdot)|_{C_0}=l_2( l_1(s_1(X^1))+\Upsilon^1_{X^1},\cdot)|_{C_0}=l_1^C\circ \Upsilon^2_{X^1}+\Upsilon^1_{X^1}\circ \varrho.
\end{eqnarray}
For  $\nabla^1$, we have
\begin{equation}\label{eq:diff3}
  \nabla^1_{\frkl_1(X^1)}=l_2(s_0\frkl_1(X^1),\cdot)|_{C_1}=l_2( l_1(s_1(X^1))+\Upsilon^1_{X^1},\cdot)|_{C_1}=\Upsilon^2_{X^1}\circ l_1^C.
\end{equation}
By \eqref{eq:diff1}, \eqref{eq:diff2} and \eqref{eq:diff3}, we deduce that
\begin{equation}\label{eq:diff}
  F^0\circ \frkl_1=\dM\circ F^1.
\end{equation}

By straightforward computation, we have
\begin{eqnarray*}
  \langle \nabla^B_{\frkl_2(X^0,Y^0)}b-\nabla^B_{X^0}\nabla^B_{Y^0}b+\nabla^B_{Y^0}\nabla^B_{X^0}b,\xi\rangle&=&\langle b,a\big(\hat{l}_2(s_0(X^0),s_0(Y_0))-s_0\frkl_2(X^0,Y^0)\big)(\xi)\rangle\\
  &=&\langle b,-a\big(R^0(X^0,Y^0)\big)(\xi)\rangle,
\end{eqnarray*}
which implies that
\begin{equation}\label{eq:morobs1}
\nabla^B_{\frkl_2(X^0,Y^0)} -\nabla^B_{X^0}\nabla^B_{Y^0} +\nabla^B_{Y^0}\nabla^B_{X^0}=\varrho\circ R^0(X^0,Y^0).
\end{equation}
Similarly, we have
\begin{eqnarray*}
 && \nabla^0_{\frkl_2(X^0,Y^0)}c^0-\nabla^0_{X^0}\nabla^0_{Y^0}c^0+\nabla^0_{Y^0}\nabla^0_{X^0}c^0 \\&=&  l_2(s_0\frkl_2(X^0,Y^0),c^0)-l_2(s_0(X^0),l_2(s_0(Y_0),c^0))+l_2(s_0(Y^0),l_2(s_0(X_0),c^0))\\
  &=& -l_1l_3(s_0(X^0),s_0(Y_0),c^0)+l_2(R^0(X^0,Y^0),c^0),
\end{eqnarray*}
which implies that
\begin{equation}\label{eq:morobs2}
 \nabla^0_{\frkl_2(X^0,Y^0)} -\nabla^0_{X^0}\nabla^0_{Y^0} +\nabla^0_{Y^0}\nabla^0_{X^0}=l_1^C\circ \Lambda(X^0,Y^0)+R^0(X^0,Y^0)\circ\varrho,
\end{equation}
and
\begin{eqnarray*}
 && \nabla^1_{\frkl_2(X^0,Y^0)}c^1-\nabla^1_{X^0}\nabla^1_{Y^0}c^1+\nabla^1_{Y^0}\nabla^1_{X^0}c^1 \\&=&  l_2(s_0\frkl_2(X^0,Y^0),c^1)-l_2(s_0(X^0),l_2(s_0(Y_0),c^1))+l_2(s_0(Y^0),l_2(s_0(X_0),c^1))\\
  &=& -l_3(s_0(X^0),s_0(Y^0),l_1(c^1))+l_2(R^0(X^0,Y^0),c^1),
\end{eqnarray*}
which implies that
\begin{equation}\label{eq:morobs3}
 \nabla^1_{\frkl_2(X^0,Y^0)} -\nabla^1_{X^0}\nabla^1_{Y^0} +\nabla^1_{Y^0}\nabla^1_{X^0}=\Lambda(X^0,Y^0)\circ l_1^C.
\end{equation}
By \eqref{eq:morobs1}, \eqref{eq:morobs2} and \eqref{eq:morobs3}, we obtain
\begin{equation}\label{eq:morobs}
  F^0(\frkl_2(X^0,Y^0))-[F^0(X^0),F^0(Y^0)]_C=\dM F^2_0(X^0,Y^0).
\end{equation}

By the equality
$$
 {l}_2(s_0(X^0), {l}_2(s_1(Y^1),c^0))+c.p.= \hat{l}_3(s_0(X^0), l_1(s_1(Y^1)),c^0),
$$
we obtain
\begin{equation}\label{eq:morobs4}
  [F^0(X^0),\Upsilon^2_{Y^1}]_C-\Upsilon^2_{\frkl_2(X^0,Y^1)}=-\Lambda(X^0,\frkl_1(Y^1))-R^1(X^0,Y^1)\circ\varrho.
\end{equation}
Furthermore, we have
\begin{eqnarray}
  \nonumber\Upsilon^1_{\frkl_2(X^0,Y^1)}&=&s_0\frkl_1(\frkl_2(X^0,Y^1))-\hat{l}_1s_1(\frkl_2(X^0,Y^1))\\
 \nonumber &=&s_0\frkl_2(X^0,\frkl_1(Y^1))-\hat{l}_1\hat{l}_2(s_0(X^0),s_1(Y^1))-\hat{l}_1R^1(X^0,Y^1)\\
  \nonumber&=&s_0\frkl_2(X^0,\frkl_1(Y^1))-\hat{l}_2(s_0(X^0),\hat{l}_1 s_1(Y^1))-{l}_1^C\circ R^1(X^0,Y^1)\\
 \nonumber &=&s_0\frkl_2(X^0,\frkl_1(Y^1))-\hat{l}_2(s_0(X^0),  s_0\frkl_1(Y^1)-\Upsilon^1_{Y^1})-{l}_1^C\circ R^1(X^0,Y^1)\\
 \label{eq:morobs5} &=&[F^0(X^0),\Upsilon^1_{Y^1}]_C+R^0(X^0,\frkl_1(Y^1))-{l}_1^C\circ R^1(X^0,Y^1).
\end{eqnarray}
By \eqref{eq:morobs4} and \eqref{eq:morobs5}, we deduce that
\begin{equation}\label{eq:morobs0}
  F^1(\frkl_2(X^0,Y^1))-[F^0(X^0),F^1(Y^1)]_C=F^2_0(X^0,\frkl_1(Y^1))-\dM F^2_1(X^0,Y^1).
\end{equation}
By straightforward computation, we have
\begin{eqnarray}
  \nonumber&&R^1(\frkl_1(X^1),Y^1)-R^1(X^1,\frkl_1(Y^1))\\
  \nonumber&=&s_1\frkl_2(\frkl_1(X^1),Y^1)-\hat{l}_2(s_0\frkl_1(X^1),s_1(Y^1))
  -s_1\frkl_2(X^1,\frkl_1(Y^1))+\hat{l}_2(s_1(X^1),s_0\frkl_1(Y^1))\\
  \nonumber&=&\hat{l}_2(s_1(X^1), \hat{l}_1 s_1(Y^1)) + \hat{l}_2(s_1(X^1), \Upsilon^1_{Y^1})-\hat{l}_2(s_0\frkl_1(X^1),s_1(Y^1))\\
  \nonumber&=& -\hat{l}_2(\Upsilon^1_{X^1},s_1({Y^1}))+ \hat{l}_2(s_1(X^1), \Upsilon^1_{Y^1})\\
  \label{eq:morobs1111}&=&[\Upsilon^1_{X^1}+\Upsilon^2_{X^1},\Upsilon^1_{Y^1}+\Upsilon^2_{Y^1}]_C.
   \end{eqnarray}

By the equality
$$
\hat{l}_2(s_0(X^0),\hat{l}_2(s_0(Y^0),s_0(Z^0)))+c.p.=\hat{l}_1\hat{l}_3(s_0(X^0), s_0(Y^0),s_0(Z^0)),
$$
we deduce that
\begin{equation}\label{eq:compF11}
  [F^0(X^0),R^0(Y^0,Z^0)]_C+R^0(X^0,\frkl_2(Y^0,Z^0))+c.p.=\Upsilon^1_{\frkl_3(X^0,Y^0,Z^0)}+l_1^C\circ \Xi(X^0, Y^0,Z^0).
\end{equation}
By the equality
$$
l_2(s_0(X^0),l_3(s_0(Y^0),s_0(Z^0),c^0))-l_3(l_2(s_0(X^0), s_0(Y^0)),s_0(Z^0),c^0)+c.p.=0,
$$
we deduce that
\begin{equation}\label{eq:compF12}
  -[F^0(X^0),\Lambda(Y^0,Z^0)]_C+\Lambda(\frkl_2(X^0,Y^0),Z^0)+c.p.+\Upsilon^2_{\frkl_3(X^0,Y^0,Z^0)}-\Xi(X^0,Y^0,Z^0)\circ\varrho=0.
\end{equation}
By \eqref{eq:compF11} and \eqref{eq:compF12}, we obtain
\begin{equation}\label{eq:compF2}
  [F^0(X^0),F^2_0(Y^0,Z^0)]_C+F^2_0(X^0,\frkl_2(Y^0,Z^0))+c.p.=F^1(\frkl_3(X^0,Y^0,Z^0))+\dM F^3(X^0,Y^0,Z^0).
\end{equation}

Then by the equality
$$
\hat{l}_2(s_0(X^0),\hat{l}_2(s_0(Y^0),s_1(Z^1)))+c.p.=\hat{l}_3(s_0(X^0), s_0(Y^0),\hat{l}_1( s_1(Z^1))),
$$
we deduce that
\begin{eqnarray}
  \nonumber&& [F^0(X^0),R^1(Y^0,Z^1)]_C+[F^0(Y^0),R^1(Z^1,X^0)]_C+[\Upsilon^2_{Z^1},R^0(X^0,Y^0)]_C\\
  \nonumber&&+R^1(X^0,\frkl_2(Y^0,Z^1))+R^1(Y^0,\frkl_2(Z^1,X^0)) +R^1(Z^1,\frkl_2(X^0,Y^0)) \\
 \label{eq:compF3} &=& \Xi(X^0, Y^0,\frkl_1(Z^1))-[\Lambda(X^0,Y^0),\Upsilon^1_{Z^1}]_C.
\end{eqnarray}

Finally, by the equality
\begin{eqnarray*}
&&\sum_{i=1}^4(-1)^{i+1}\hat{l}_2\big(s_0(X^0_i),\hat{l}_3(s_0(X^0_1),\cdots,\widehat{s_0(X^0_i)},\cdots,s_0(X^0_4))\big)\\
&&+\sum_{i<j,k<l}(-1)^{i+j}\hat{l}_3\big(\hat{l}_2(s_0(X^0_i),s_0(X^0_j)),s_0(X^0_k),s_0(X^0_l)\big)=0,
\end{eqnarray*}
we deduce that
\begin{eqnarray}
\nonumber&&\sum_{i=1}^4(-1)^{i+1}\Big([F^0(X^0_i),\Xi(X^0_1,\cdots,\widehat{X^0_i},\cdots, X^0_4)]_C+R^1(X^0_i,\frkl_3(X^0_1,\cdots,\widehat{X^0_i},\cdots, X^0_4))\Big)\\
\nonumber&&+\sum_{i<j}(-1)^{i+j}\Big(\Xi(\frkl_2(X^0_i,X^0_j),X^0_1,\cdots,\widehat{X^0_i},\cdots,\widehat{X^0_j},\cdots, X^0_4)\\
 \label{eq:compF4}&&-[R^0(X^0_i,X^0_j),\Lambda(X^0_1,\cdots,\widehat{X^0_i},\cdots,\widehat{X^0_j},\cdots, X^0_4)]_C\Big)=0.
\end{eqnarray}

By \eqref{eq:anccom}, \eqref{eq:diff}, \eqref{eq:morobs}, \eqref{eq:morobs0}, \eqref{eq:morobs1111}, \eqref{eq:compF2}, \eqref{eq:compF3} and \eqref{eq:compF4}, we deduce that $(F^0,F^1,F^2_0,F^2_1,F^3)$ is a morphism from the Lie 2-algebroid $(A_{-1},A_0,\frka,\frkl_1,\frkl_2,\frkl_3)$ to the strict Lie 3-algebroid $$(\End^{-2}(\huaE),\End^{-1}(\huaE),\dev_\pi(\huaE),\frkp,\dM,[\cdot,\cdot]_C).$$

Conversely, let $(A_{-1}, A_0, \frka, \frkl_1, \frkl_2, \frkl_3)$  be  a  Lie 2-algebroid and  $(F^0,F^1,F^2_0,F^2_1,F^3)$ a  flat superconnection  on the $3$-term complex $C_{-1}\stackrel{l_1^C}{\longrightarrow}C_{0}\stackrel{\varrho}{\longrightarrow}B$. Then we can obtain a $\VB$-Lie $2$-algebroid structure on the split graded double vector bundle  $\left(\begin{array}{ccc}
A_{-1}\oplus B\oplus C_{-1};&A_{-1},B;&M\\
A_0\oplus B\oplus C_0;&A_{0},B;&M
\end{array}\right).$
 We leave the details to readers. The proof is finished. \qed

\section{$\VB$-$\LWX$ 2-algebroids}

In this section, first we recall the notion of a $\LWX$ 2-algebroid, and then
  explore what is a   metric graded double vector bundle,  introduce the notion of a $\VB$-$\LWX$ 2-algebroid, which can be viewed as the categorification of a $\VB$-Courant algebroid introduced in \cite{Li-Bland}.

As a  model for ``Leibniz
algebras that satisfy Jacobi identity up to all higher homotopies'',
  the notion of a strongly homotopy
Leibniz algebra, or a $Lod_\infty$-algebra was given in \cite{livernet} by Livernet,
 which was further studied by Ammar and Poncin
in \cite{ammardefiLeibnizalgebra}.
  In \cite{Leibniz2al}, the authors introduced the notion of a Leibniz 2-algebra, which is the categorification of a Leibniz algebra, and proved that the category of Leibniz 2-algebras and the category of 2-term $Lod_\infty$-algebras are equivalent. Due to this reason, a 2-term $Lod_\infty$-algebra will be called a Leibniz 2-algebra directly in the sequel.

\begin{defi}{\rm(\cite{LiuSheng})}\label{defi:Courant-2 algebroid}
A {\bf $\LWX$ $2$-algebroid} is a graded vector bundle $\huaE=E_{-1}\oplus E_0$ over $M$ equipped with a non-degenerate graded symmetric bilinear form  $S$ on $\huaE$, a bilinear operation $\diamond:\Gamma(E_{-i})\times \Gamma(E_{-j})\longrightarrow \Gamma(E_{-(i+j)})$, $0\leq i+j\leq 1$, which is skewsymmetric  on $\Gamma(E_0)\times \Gamma(E_0)$, an $E_{-1}$-valued $3$-form $\Omega$ on $E_0$, two bundle maps $\partial:E_{-1}\longrightarrow E_0$ and $\rho:E_0\longrightarrow TM$, such that $E_{-1}$ and $E_0$ are isotropic and the following axioms are satisfied:
\begin{itemize}
\item[$\rm(i)$]$(\Gamma(E_{-1}),\Gamma(E_0),\partial,\diamond,\Omega)$ is a Leibniz $2$-algebra;
\item[$\rm(ii)$]for all $e\in\Gamma(\huaE)$, $e\diamond e=\frac{1}{2}\huaD S(e,e)$, where $\huaD:\CWM\longrightarrow \Gamma(E_{-1})$ is defined by
 \begin{equation}
 S(\huaD f,e^0)=\rho(e^0)(f),\quad \forall f\in\CWM, ~e^0\in\Gamma(E_0);
 \end{equation}
\item[$\rm(iii)$]for all $e^1_1,e^1_2\in\Gamma(E_{-1})$, $S( \partial(e^1_1),e^1_2)=S(e^1_1,\partial(e^1_2))$;
\item[$\rm(iv)$]for all $e_1,e_2,e_3\in\Gamma(\huaE)$, $\rho(e_1)S( e_2,e_3)=S( e_1\diamond e_2,e_3)+S(e_2,e_1\diamond e_3)$;
\item[$\rm(v)$]for all $e^0_1,e^0_2,e^0_3,e^0_4\in\Gamma(E_0)$, $S(\Omega(e^0_1,e^0_2,e^0_3),e^0_4)=-S(e^0_3,\Omega(e^0_1,e^0_2,e^0_4))$.
 \end{itemize}
\end{defi}
 Denote a $\LWX$ 2-algebroid  by $(E_{-1},E_0,\partial,\rho,S,\diamond,\Omega)$, or simply by $\huaE$. Since the section space of a $\LWX$ 2-algebroid is a Leibniz 2-algebra, the section space of a Courant algebroid is a Leibniz algebra and  Leibniz 2-algebras are the categorification of Leibniz algebras, we can view $\LWX$ 2-algebroids   as the categorification of Courant algebroids.

 As a higher analogue of Roytenberg's result about symplectic NQ manifolds of degree 2 and Courant algebroids (\cite{royt}), we have
 \begin{thm}{\rm(\cite{LiuSheng})}\label{thm:QPC2A}
  Let $(T^*[3]A^*[2],\Theta)$ be a symplectic NQ manifold of degree $3$, where $A$ is an ordinary vector bundle and $\Theta$ is a degree $4$ function on $T^*[3]A^*[2]$ satisfying $\{\Theta,\Theta\}=0$. Here $\{\cdot,\cdot\}$ is the canonical Poisson bracket on $T^*[3]A^*[2]$. Then $( A^*[1],A,\partial, \rho,S,\diamond, \Omega )$ is  a $\LWX$ $2$-algebroid, where the bilinear form  $S$ is given by
  $$
  S(X+\alpha,Y+\beta)=\langle X,\beta\rangle+\langle Y,\alpha \rangle,\quad \forall~X,Y\in\Gamma(A),\alpha,\beta\in\Gamma(A^*),
  $$
 and $\partial$, $\rho$,   $\diamond$  and  $\Omega$ are given by derived brackets. More precisely, we have
  \begin{eqnarray*}
\partial\alpha&=&\{\alpha,\Theta\},\quad \forall~\alpha\in\Gamma(A^*),\label{eq:Q5}\\
\rho(X)(f)&=&\{f,\{X,\Theta\}\},\quad \forall~X\in\Gamma(A),f\in\CWM,\label{eq:Q4}\\
X\diamond Y&=&\{Y,\{X,\Theta\}\},\quad\forall~ X,Y\in\Gamma(A),\\
X\diamond\alpha&=&\{\alpha,\{X,\Theta\}\},\quad\forall~ X\in\Gamma(A),\alpha\in\Gamma(A^*),\\
\alpha\diamond X&=&-\{X,\{\alpha,\Theta\}\},\quad\forall~ X\in\Gamma(A),\alpha\in\Gamma(A^*),\\
\Omega(X,Y,Z)&=&\{Z,\{Y,\{X,\Theta\}\}\},\quad \forall~X,Y,Z\in\Gamma(A).\label{eq:Q7}
\end{eqnarray*}
\end{thm}

See \cite{Kosmann-Schwarzbach,Voronov1} for more information of derived brackets.

  \begin{defi}
 A {\bf metric graded double vector bundle} is a graded double vector bundle $\left(\begin{array}{ccc}
D_{-1};&A_{-1},B;&M\\
D_{0};&A_{0},B;&M
\end{array}\right)$ equipped with a  degree $1$ nondegenerate graded symmetric bilinear form $S$ on the graded bundle $D^B_{-1}\oplus D^B_0$ such that   it induces an isomorphism between graded double vector bundles\begin{equation*}
\xymatrix@!0{  D_{-1} \ar@{->} [rr]   \ar@{->} [rd]
&&  B   \ar@{->} [rd]&&&\\
&A_{-1}\ar@{->} [rr]&&M &&C_{-1}\ar@{->}[ll]\\
D_0\ar@{->} [rr] \ar@{->} [rd]&&B\ar@{->} [rd]& &&\\
&A_0\ar@{->} [rr]&&M&&C_0\ar@{->}[ll]}\mbox{\rm and}\qquad
 \xymatrix@!0{  D_0^{\star B}[1]\ar@{->} [rr]   \ar@{->} [rd]
&&  B   \ar@{->} [rd]&&&\\
&C_{0}^*\ar@{->} [rr] &&M &&A_{0}^*\ar@{->}[ll]\\
D_{-1}^{\star B}[1]\ar@{->} [rr] \ar@{->} [rd]&&B\ar@{->} [rd]& &&\\
&C_{-1}^*\ar@{->} [rr]&&M&&A_{-1}^*,\ar@{->}[ll]}
\end{equation*}
where $\star B$ means dual over $B$.
  \end{defi}

Given a metric graded double vector bundle, we have
$$
C_0\cong A_{-1}^*,\quad C_{-1}\cong A_0^*.
$$
    In the sequel, we will always identify $C_0$ with $ A_{-1}^*$, $ C_{-1}$ with $ A_0^*$. Thus, a metric graded double vector bundle is of the following form:
        \begin{equation*}
\xymatrix@!0{  D_{-1} \ar@{->} [rr]   \ar@{->} [rd]
&&  B    \ar@{->} [rd]&&&\\
&A_{-1}\ar@{->} [rr] &&M &&A_0^*\ar@{->}[ll]\\
D_0\ar@{->} [rr] \ar@{->} [rd]&&B \ar@{->} [rd]& &&\\
&A_0\ar@{->} [rr]&&M& &A_{-1}^*.\ar@{->}[ll]}
\end{equation*}

 Now we are ready to put a $\LWX$ 2-algebroid structure on a graded double vector bundle.
\begin{defi}
A {\bf $\VB$-$\LWX$ 2-algebroid} is a metric graded double vector bundle $$\left(\left(\begin{array}{ccc}
D_{-1};&A_{-1},B;&M\\
D_{0};&A_{0},B;&M
\end{array}\right),S\right),$$ equipped with a $\LWX$ $2$-algebroid structure $(D_{-1}^B,D_0^B,\partial,\rho,S,\diamond,\Omega)$ such that
\begin{itemize}
  \item[\rm(i)]  $\partial $ is linear, i.e. there exists a unique bundle map $\overline{\partial}:A_{-1}\longrightarrow A_0$ such that $\partial: D_{-1}\longrightarrow D_0$ is a double vector bundle morphism over $\overline{\partial}:A_{-1}\longrightarrow A_0$ (see Diagram {\rm(iii)});

  \item[\rm(ii)]  the anchor $\rho$ is a linear, i.e. there exists a unique bundle map $\overline{\rho}:A_{0}\longrightarrow TM$  such that $\rho:D_0\longrightarrow TB$ is a  double vector bundle morphism over $\overline{\rho}:A_{0}\longrightarrow TM$ (see Diagram {\rm(iv)});
   \begin{equation*}
\xymatrix@!0{  D_{-1} \ar@{->} [rr] \ar@ {->} [dd]_{\partial} \ar@{->} [rd]
&&  B  \ar@{=}[dd] \ar@{->} [rd]&&&\\
&A_{-1}\ar@{->} [rr]\ar@{->} [dd]_{\overline{\partial}}&&M\ar@{=} [dd]&&A_{0}^*\ar@{->}[ll]\\
D_0 \ar@{->} [rd]&\ar@{->} [r]&B \ar@{->} [rd]& &&\\
&A_0\ar@{->} [rr]&&M&& A_{-1}^*\ar@{->}[ll]\\
&&\mbox{\rm Diagram (iii)}}
\qquad
\xymatrix@!0{  D_0 \ar@{->} [rr] \ar@ {->} [dd]_{\rho } \ar@{->} [rd]
&&  B  \ar@{=}[dd] \ar@{->} [rd]&&&\\
&A_0\ar@{->} [rr]\ar@{->} [dd]_{\overline{\rho}}&&M\ar@{=} [dd]&&A_{-1}^*\ar@{->}[ll]\\
TB \ar@{->} [rd]&\ar@{->} [r]&B \ar@{->} [rd]& &&\\
&TM\ar@{->} [rr]&&M&& B,\ar@{->}[ll]\\
&&\mbox{\rm Diagram (iv)}}
 \end{equation*}
   \item[\rm(iii)] the   operation $\diamond$ is linear;
     \item[\rm(iv)] $\Omega$ is linear.
 \end{itemize}
\end{defi}

Since a $\LWX$ $2$-algebroid can be viewed as the categorification of a Courant algebroid, we can view a $\VB$-$\LWX$ $2$-algebroid as the categorification of a $\VB$-Courant algebroid.

\begin{ex}{\rm
  Let $(A_{-1},A_0,a,l_1,l_2,l_3)$
 be a Lie 2-algebroid. Let $E_0=A_0\oplus A^*_{-1}$, $E_{-1}=A_{-1}\oplus A^*_{0}$ and $\huaE=E_0\oplus E_{-1}$.
 Then $(E_{-1},E_0,\partial,\rho,S,\diamond,\Omega)$ is a $\LWX$ $2$-algebroid, where $\partial:E_{-1}\longrightarrow E_0$ is given by
  $$
  \partial(X^1+\alpha^0)=l_1(X^1)+l^*_1(\alpha^0),\quad \forall X^1\in\Gamma(A_{-1}),~\alpha^0\in\Gamma(A^*_0),
  $$
  $\rho:E_0\longrightarrow TM$ is given by
   $$
   \rho(X^0+\alpha^1)=a(X^0),\quad \forall X^0\in\Gamma(A_0),~ \alpha^1\in\Gamma(A_{-1}^*),
   $$ the symmetric bilinear form $S=(\cdot,\cdot)_+$ is given by
    \begin{equation} \label{eq:naturalsymform}
(X^0+\alpha^1+X^1+\alpha^0,Y^0+\beta^1+Y^1+\beta^0)_+=\langle X^0,\beta^0 \rangle+\langle Y^0,\alpha^0 \rangle+\langle X^1,\beta^1 \rangle+\langle Y^1,\alpha^1 \rangle,
\end{equation}
the operation $\diamond$ is given by
\begin{equation}\label{eq:0-bracket}\left\{\begin{array}{rcl}
(X^0+\alpha^1)\diamond(Y^0+\beta^1)&=&l_2(X^0,Y^0)+L^0_{X^0}\beta^1-L^0_{Y^0}\alpha^1,\\
(X^0+\alpha^1)\diamond(X^1+\alpha^0)&=&l_2(X^0,X^1)+L^0_{X^0}\alpha^0+\iota_{X^1}\delta(\alpha^1), \\
(X^1+\alpha^0)\diamond(X^0+\alpha^1)&=&l_2(X^1,X^0)+L^1_{X^1}\alpha^1-\iota_{X^0}\delta(\alpha^0),
\end{array}\right.
\end{equation} and the $E_{-1}$-valued $3$-form $\Omega$ is defined by
\begin{equation}\label{eq:st3-form}
\Omega(X^0+\alpha^1,Y^0+\beta^1,Z^0+\zeta^1)=l_3(X^0,Y^0,Z^0)+L^3_{X^0,Y^0}\zeta^1+L^3_{Z^0,X^0}\beta^1+L^3_{Y^0,Z^0}\alpha^1,
\end{equation}
where $L^0,L^1,L^3$ are given by \eqref{eq:L}. It is straightforward to see that this $\LWX$ 2-algebroid  gives rise to a $\VB$-$\LWX$ 2-algebroid:
   \begin{equation*}
\xymatrix@!0{ A_{-1}\times_M A_0^*  \ar@{->} [rr] \ar@ {->} [dd] \ar@{->} [rd]
&& M\ar@ {=} [dd] \ar@{->} [rd]&\\
& A_{-1}\ar@{->} [rr]\ar@{->} [dd]&&M\ar@{=} [dd]\\
A_0\times_M A_{-1}^*\ar@{->} [rr] \ar@{->} [rd]&&M\ar@{->} [rd]& \\
& A_0\ar@{->} [rr]&&M.}
\end{equation*}
}
\end{ex}

\begin{ex}\label{ex:TE}{\rm
  For any manifold $M$, $(T^*[1]M,TM,\partial=0,\rho={\id}, S,\diamond,\Omega=0)$ is a $\LWX$ 2-algebroid, where $S$ is the natural symmetric pairing between $TM$ and $T^*M$, and $\diamond$ is the standard Dorfman bracket given by
  \begin{equation}
  (X+\alpha)\diamond(Y+\beta)=[X,Y]+L_X\beta-\iota_Y d\alpha,\quad\forall~X,Y\in\frkX(M),~\alpha,\beta\in\Omega^1(M).
  \end{equation}
See \cite[Remark 3.4]{LiuSheng} for more details.  In particular, for any vector bundle $E$, $(T^*E^*,TE^*,\partial=0,\rho={\id}, S,\diamond,\Omega=0)$ is a $\LWX$ 2-algebroid, which gives rise to a $\VB$-$\LWX$ 2-algebroid:
  \begin{equation*}
\xymatrix@!0{ T^*E^* \ar@{->} [rr] \ar@ {->} [dd] \ar@{->} [rd]
&& E^*\ar@ {=} [dd] \ar@{->} [rd]&&&\\
& E\ar@{->} [rr]\ar@{->} [dd]&&M\ar@{=} [dd]&&T^*M\ar@{->} [ll]\\
TE^*\ar'[r] [rr] \ar@{->} [rd]&&E^*\ar@{->} [rd]&& &\\
& TM\ar@{->} [rr]&&M&&E^*.\ar@{->} [ll]}
\end{equation*}
  }
\end{ex}

We have a higher analogue of  Theorem \ref{thm:correspondence}:
\begin{thm}\label{thm:Lie3-VBLWX}
  There is a one-to-one correspondence between split Lie $3$-algebroids and split $\VB$-$\LWX$ $2$-algebroids.
\end{thm}

\pf  Let $\huaA=(A_{-2},A_{-1},A_0,a,l_1,l_2,l_3,l_4)$ be a split Lie 3-algebroid. Then $T^*[3]\huaA[1]$ is a symplectic NQ manifold of degree 3. Note that
$$ T^*[3]\huaA[1]=T^*[3]( A_0\times_MA_{-1}^*\times_M A_{-2}^*)[1],$$ where $ A_0\times_MA_{-1}^*\times_M A_{-2}^*$ is viewed as a vector bundle over the base $A_{-2}^*$ and $A_{-1}\times_M A_{0}^*\times_M A_{-2}^*$ is its dual bundle. Denote by $(x^i,\mu_j,\xi^k,\theta_l,p_i,\mu^j,\xi_k,\theta^l)$ a canonical (Darboux) coordinate on $T^*[3](A_0\times_MA_{-1}^*\times_M A_{-2}^*)[1]$, where $x^i$ is a smooth coordinate on $M$, $\mu_j\in\Gamma(A_{-2})$ is a fibre coordinate on $A_{-2}^*$, $\xi^k\in\Gamma(A_{0}^*)$ is a fibre coordinate on $A_{0}$,  $\theta_l\in\Gamma(A_{-1})$ is a fibre coordinate on $A_{-1}^*$ and $(p_i,\mu^j,\xi_k,\theta^l)$ are the momentum coordinates for $(x^i,\mu_j,\xi^k,\theta_l)$. About their degrees, we have
$$
\left(\begin{array}{cccccccc}
x^i&\mu_j&\xi^k&\theta_l&p_i&\mu^j&\xi_k&\theta^l\\
0&0&1&1&3&3&2&2
\end{array}
\right)
$$
The symplectic structure is given by
$$
\omega=dx^idp_i+d\mu_jd\mu^j+d\xi^kd\xi_k+d\theta_ld\theta^l,
$$
which is degree 3. The Lie 3-algebroid structure gives rise to a degree 4 function $\Theta$ satisfying $\{\Theta,\Theta\}=0.$ By Theorem \ref{thm:QPC2A}, we obtain a $\LWX$ 2-algebroid $(D_{-1},D_0,\partial,\rho,S,\diamond,\Omega)$, where $D_{-1}=A_{-1}\times_M A_0^* \times_M A_{-2}^*$ and $D_0= A_0\times_MA_{-1}^*\times_M A_{-2}^*$ are vector bundles over $A_{-2}^*$. Obviously, they give the graded double vector bundle
$$ \left(\begin{array}{ccc}
A_{-1}\times_M A_0^* \times_M A_{-2}^*;&A_{-1},A_{-2}^*;&M\\
A_0\times_MA_{-1}^*\times_M A_{-2}^*;&A_{0},A_{-2}^*;&M
\end{array}\right). $$
   The section space $\Gamma_{A_{-2}^*}(D_0)$ are generated by $\Gamma(A_{-1}^*)$ (the space of core sections) and $\Gamma(A_{-2}\otimes A_{-1}^*)\oplus \Gamma(A_0)$ (the space of linear sections) as $C^\infty(A_{-2}^*)$-module. Similarly, The section space $\Gamma_{A_{-2}^*}(D_{-1})$ are generated by $\Gamma(A_{0}^*)$  and $\Gamma(A_{-2}\otimes A_{0}^*)\oplus \Gamma(A_{-1})$  as $C^\infty(A_{-2}^*)$-module. Thus, in the sequel we only consider core sections and linear sections.

  The graded symmetric bilinear form $S$ is given by
  \begin{equation}
   S(e^0,e^1)= S(X^0+\psi^1+\alpha^1,X^1+\psi^0+\alpha^0)=\langle \alpha_1,X^1\rangle+\langle \alpha^0,X_0\rangle+\psi^1(X^1)+\psi^0(X^0),
  \end{equation}
  for all $e^0=X^0+\psi^1+\alpha^1\in\Gamma_{A_{-2}^*}(D_0)$ and $e^1=X^1+\psi^0+\alpha^0\in\Gamma_{A_{-2}^*}(D_{-1})$, where $X^i\in\Gamma(A_{-i})$, $\psi^i\in\Gamma( A_{-2}\otimes A_{-i}^*)$ 
   and $\alpha^i\in\Gamma(A_{-i}^*)$. Then it is obvious that
  $$\left(\left(\begin{array}{ccc}
A_{-1}\times_M A_0^* \times_M A_{-2}^*;&A_{-1},A_{-2}^*;&M\\
A_0\times_MA_{-1}^*\times_M A_{-2}^*;&A_{0},A_{-2}^*;&M
\end{array}\right),S\right)$$
  is a metric graded double vector bundle.

  The bundle map $\partial:D_{-1}\longrightarrow D_0$ is given by
  $$
  \partial(X^1+\psi^0+\alpha^0)=l_1(X^1)+l_2(X^1,\cdot)|_{A_{-1}}+\psi^0\circ l_1+l_1^*(\alpha^0).
  $$
Thus, $\partial:D_{-1}\longrightarrow D_0$ is a double vector bundle morphism over $l_1:A_{-1}\longrightarrow A_0$.

Note that functions on $A_{-2}^*$ are generated by fibrewise constant functions $\CWM$ and fibrewise linear functions $\Gamma(A_{-2})$. For all $f\in\CWM$ and $X^2\in\Gamma(A_{-2})$, the anchor $\rho:D_0\longrightarrow TA_{-2}^*$ is given by
$$
\rho(X^0+\psi^1+\alpha^1)(f+X^2)=a(X^0)(f)+\langle \alpha^1,l_1(X^2)\rangle+l_2(X^0,X^2)+\psi^1 (l_1(X^2)).
$$
  Therefore, for a linear section $X^0+\psi^1\in\Gamma^l_{A_{-2}^*}(D_0)$, the image $\rho(X^0+\psi^1)$ is a linear vector field and for a core section $\alpha^1\in\Gamma(A_{-1}^*)$, the image $\rho(\alpha^1)$ is a constant vector field. Thus, $\rho$ is linear.

   The bracket operation $\diamond$ is given by
   \begin{eqnarray*}
    &&( X^0+\psi^1+\alpha^1)\diamond (Y^0+\phi^1+\beta^1)\\
    &=&l_2(X^0,Y^0)+l_3(X^0,Y^0,\cdot)|_{A_{-1}}+l_2(X^0,\phi^1(\cdot))-\phi^1\circ l_2(X^0,\cdot)|_{A_{-1}}+L^0_{X_0}\beta^1\\
    &&+\psi^1\circ l_2(Y^0,\cdot)|_{A_{-1}}-l_2(Y^0,\psi^1(\cdot))+\psi^1\circ l_1\circ \phi^1-\phi^1\circ l_1\circ \psi^1-\beta^1\circ l_1\circ\psi^1\\
    &&-L^0_{Y_0}\alpha^1+\alpha^1\circ l_1\circ \phi^1,\\
     &&( X^0+\psi^1+\alpha^1)\diamond (Y^1+\phi^0+\beta^0)\\
     &=&l_2(X^0,Y^1)+l_3(X^0,\cdot,Y^1)|_{A_{0}}+l_2(X^0,\phi^0(\cdot))-\phi^0\circ l_2(X^0,\cdot)|_{A_{0}}+L^0_{X^0}\beta^0\\
     &&-\psi^1l_2(\cdot,Y^1)|_{A_0}+\delta (\psi^1(Y^1))+\psi^1\circ l_1\circ \phi^0+\iota_{Y_1}\delta\alpha^1+\alpha^1\circ l_1\circ\phi^0,\\
     &&(Y^1+\phi^0+\beta^0)\diamond( X^0+\psi^1+\alpha^1)\\
     &=&l_2(Y^1,X^0)-l_3(X^0,\cdot,Y^1)|_{A_{0}}-l_2(X^0,\phi^0(\cdot))+\phi^0\circ l_2(X^0,\cdot)|_{A_{0}}+\delta(\phi^0(X^0))\\&&-\iota_{X^0}\delta\beta^0
+\psi^1l_2(\cdot,Y^1)|_{A_0}-\psi^1\circ l_1\circ \phi^0+L^1_{Y_1}\alpha^1-\alpha^1\circ l_1\circ\phi^0.
   \end{eqnarray*}
   Then it is straightforward to see that the operation $\diamond$ is linear.

  Finally, $\Omega$ is given by
\begin{eqnarray*}
  &&\Omega(X^0+\psi^1+\alpha^1,Y^0+\phi^1+\beta^1,Z^0+\varphi^1+\gamma^1)\\
  &=&l_3(X^0,Y^0,Z^0)+l_4(X^0,Y^0,Z^0,\cdot)\\
  &&-\varphi^1\circ l_3(X^0,Y^0,\cdot)|_{A_0}-\phi^1\circ l_3(Z^0,X^0,\cdot)|_{A_0}-\psi^1\circ l_3(Y^0,Z^0,\cdot)|_{A_0}\\
  &&+L^3_{X^0,Y^0}\gamma^1+L^3_{Y^0,Z^0}\alpha^1+L^3_{Z^0,X^0}\beta^1,
\end{eqnarray*}
which implies that $\Omega$ is also linear.

Thus, a split Lie 3-algebroid gives rise to a split $\VB$-$\LWX$ 2-algebroid:
   \begin{equation*}
\xymatrix{  A_{-1}\times_M A_0^* \times_M A_{-2}^* \ar@{->} [rr] \ar@ {->} [dd]_{\partial} \ar@{->} [rd]
&& A_{-2}^* \ar@ {=} [dd] \ar@{->} [rd]&&&\\
&A_{-1}\ar@{->} [rr]\ar@{->} [dd]_{l_1}&&M\ar@{=} [dd]&&A_0^*\ar@{->} [ll]\\
 A_0\times_M A_{-1}^* \times_M A_{-2}^*\ar'[r] [rr] \ar@{->} [rd]&&A_{-2}^*\ar@{->} [rd]&&& \\
&A_0\ar@{->} [rr]&&M&&A_{-1}^*.\ar@{->} [ll]}
\end{equation*}

Conversely, given a split $\VB$-$\LWX$ 2-algebroid:
 \begin{equation*}
\xymatrix{  A_{-1}\times_M A_0^* \times_M B \ar@{->} [rr] \ar@ {->} [dd]_{\partial} \ar@{->} [rd]
&& B \ar@ {=} [dd] \ar@{->} [rd]&&&\\
&A_{-1}\ar@{->} [rr]\ar@{->} [dd]_{l_1}&&M\ar@{=} [dd]&&A_0^*\ar@{->} [ll]\\
 A_0\times_M A_{-1}^* \times_M B\ar'[r] [rr] \ar@{->} [rd]&&B\ar@{->} [rd]&&& \\
&A_0\ar@{->} [rr]&&M&&A_{-1}^*,\ar@{->} [ll]}
\end{equation*}
then we can deduce that the corresponding symplectic NQ-manifold of degree 3 is $T^*[3]\huaA[1]$, where $\huaA=A_0\oplus A_{-1}\oplus B$ is a graded vector bundle in which $B$ is of degree $-2$, and the $Q$-structure   gives rise to a Lie $3$-algebroid structure on $\huaA$. We omit details.
\qed

\begin{rmk}
   Since every double vector bundle is splitable, every $\VB$-$\LWX$ $2$-algebroid is isomorphic to a split one. Meanwhile, by choosing a splitting, we obtain a  split Lie $3$-algebroid from an NQ-manifold of degree $3$ (Lie $3$-algebroid).  Thus, we can enhance the above result to be a one-to-one correspondence between Lie $3$-algebroids and $\VB$-$\LWX$ $2$-algebroids. We omit such details.
\end{rmk}

Recall that the tangent prolongation of a Courant algebroid is a $\VB$-Courant algebroid (\cite[Proposition 3.4.1]{Li-Bland}). Now we show that the tangent prolongation of a $\LWX$ 2-algebroid is a $\VB$-$\LWX$ 2-algebroid. The notations used below is the same as the ones used in Section 3.

\begin{pro}\label{pro:TLWX2}
  Let  $(E_{-1},E_0,\partial,\rho,S,\diamond,\Omega)$ be a $\LWX$ $2$-algebroid. Then $(TE_{-1}, TE_0,\widetilde{\partial},\widetilde{\rho},\widetilde{S},\widetilde{\diamond},\widetilde{\Omega})$ is a $\LWX$ $2$-algebroid over $TM$, where the bundle map $\widetilde{\partial}:TE_{-1}\longrightarrow TE_0$ is given by
  \begin{equation}
    \widetilde{\partial}(\sigma^1_T)=\partial(\sigma^1)_T,\quad  \widetilde{\partial}(\sigma^1_C)=\partial(\sigma^1)_C,
  \end{equation}
  the bundle map $\widetilde{\rho}:TE_0\longrightarrow TTM$ is given by
  \begin{equation}
    \widetilde{\rho}(\sigma^0_T)=\rho(\sigma^0)_T,\quad  \widetilde{\rho}(\sigma^0_C)=\rho(\sigma^0)_C,
  \end{equation}
  the degree $1$ bilinear form  $\widetilde{S}$ is given by
  \begin{equation}
    \widetilde{S}(\sigma^0_T,\tau^1_T)=S(\sigma^0,\tau^1)_T,~  \widetilde{S}(\sigma^0_T,\tau^1_C)=S(\sigma^0,\tau^1)_C,~ \widetilde{S}(\sigma^0_C,\tau^1_T)=S(\sigma^0,\tau^1)_C,~ \widetilde{S}(\sigma^0_C,\tau^1_C)=0,
  \end{equation}
    the   bilinear operation  $\widetilde{\diamond}$ is given by
  \begin{eqnarray*}
  &&\sigma^0_T \widetilde{\diamond}\tau^0_T = (\sigma^0 {\diamond}\tau^0)_T,\quad   \sigma^0_T \widetilde{\diamond} \tau^0_C =-\tau^0_C\widetilde{\diamond}\sigma^0_T =(\sigma^0\diamond\tau^0)_C,\quad \sigma^0_C \widetilde{\diamond} \tau^0_C=0,\\
  &&\sigma^0_T \widetilde{\diamond}\tau^1_T = (\sigma^0 {\diamond}\tau^1)_T,\quad   \sigma^0_T \widetilde{\diamond} \tau^1_C=\sigma^0_C \widetilde{\diamond} \tau^1_T=(\sigma^0\diamond\tau^1)_C, \quad
  \sigma^0_C \widetilde{\diamond} \tau^1_C=0,\\
  &&\tau^1_T\widetilde{\diamond}\sigma^0_T =(\tau^1\diamond\sigma^0)_T,\quad\tau^1_C\widetilde{\diamond}\sigma^0_T=\tau^1_T\widetilde{\diamond}\sigma^0_C =(\tau^1\diamond\sigma^0)_C,\quad \tau^1_C\widetilde{\diamond}\sigma^0_C =0,
  \end{eqnarray*}
  and $\widetilde{\Omega}:\wedge^3TE_0\longrightarrow TE_{-1}$ is given by
  \begin{eqnarray*}
   \widetilde{\Omega}(\sigma^0_T,\tau^0_T,\varsigma^0_T)=\Omega(\sigma^0,\tau^0,\varsigma^0)_T,\quad \widetilde{\Omega}(\sigma^0_T,\tau^0_T,\varsigma^0_C)=\Omega(\sigma^0,\tau^0,\varsigma^0)_C,\quad
   \widetilde{\Omega}(\sigma^0_T,\tau^0_C,\varsigma^0_C)=0,
  \end{eqnarray*}
  for all $\sigma^0,\tau^0,\varsigma^0\in \Gamma(E_0)$ and $\sigma^1,\tau^1\in \Gamma(E_{-1})$.

 Moreover, we have the following  $\VB$-$\LWX$ $2$-algebroid:
  \begin{equation*}
\xymatrix@!0{ TE_{-1} \ar@{->} [rr] \ar@ {->} [dd] \ar@{->} [rd]
&& TM\ar@ {->} [dd] \ar@{->} [rd]&&&\\
& E_{-1}\ar@{->} [rr]\ar@{->} [dd]&&M\ar@{->} [dd]&&E_{-1}\ar@{->} [ll]\\
TE_0\ar'[r] [rr] \ar@{->} [rd]&&TM\ar@{->} [rd]&& &\\
& E_0\ar@{->} [rr]&&M&&E_0.\ar@{->} [ll]}
\end{equation*}
\end{pro}
\pf  Since  $(E_{-1},E_0,\partial,\rho,S,\diamond,\Omega)$ is a $\LWX$ $2$-algebroid, we deduce that $(TE_{-1}, TE_0,\widetilde{\partial},\widetilde{\rho},\widetilde{S},\widetilde{\diamond},\widetilde{\Omega})$ is a $\LWX$ $2$-algebroid over $TM$. Moveover, it is obvious that $\widetilde{\partial},\widetilde{\rho},\widetilde{S},\widetilde{\diamond},\widetilde{\Omega}$ are all linear, which implies that we have a $\VB$-$\LWX$ $2$-algebroid.\qed \vspace{3mm}

At the end of this section, we give the definition of a $\VB$-Lie $2$-bialgebroid and show that the double of a $\VB$-Lie $2$-bialgebroid is a $\VB$-$\LWX$ $2$-algebroid. Let us recall the notion of a split Lie 2-bialgebroid briefly. A split Lie $2$-algebroid can be described by a degree $4$ function $\mu$ on $T^*[3]\huaA[1]=T^*[3](A_0\oplus A_{-1}^*)[1]$ satisfying $\{\mu,\mu\}=0,$ where $\{\cdot,\cdot\}$ is the canonical graded Poisson bracket on $T^*[3](A_0\oplus A_{-1}^*)[1]$.
Denote by $(x^i,\xi^j,\theta_k,p_i,\xi_j,\theta^k)$ a canonical Darboux coordinate on $T^*[3](A_0\oplus A_{-1}^*)[1]$, where $x^i$ is a coordinate on $M$, $(\xi^j,\theta_k)$ is the fiber coordinate on $A_0\oplus A^*_{-1}$, $(p_i,\xi_j,\theta^k)$ is the momentum coordinate on $\huaM$ for $(x^i,\xi^j,\theta_k)$. The degree of the symplectic structure $\omega=dx^idp_i+d\xi^j d\xi_j+d\theta_kd\theta^k$ is $3$ and the degree  of the corresponding graded Poisson structure is $-3$. The degree 4 function $\mu$ is given by
\begin{equation} \label{eq:L2Amu}
\mu={\mu_1}^i_j(x)p_i\xi^j+{\mu_2}^{i}_j(x)\xi_i\theta^j+\half{\mu_3}^k_{ij}(x)\xi_k\xi^i\xi^j+ {\mu_4}_{ij}^k\theta^j\xi^i\theta_k+\frac{1}{6}{\mu_5}_{ijk}^l(x)\theta_l\xi^i\xi^j\xi^k.
\end{equation}
We will write $\mu_2={\mu_2}^{i}_j(x)\xi_i\theta^j$.

\begin{defi}{\rm(\cite{LiuSheng})}
Let $\huaA=(A_{-1},A_0,a,l_1,l_2,l_3)$ and $\huaA^*[1]=(A_0^*[1],A_{-1}^*[1],\frka,\frkl_1,\frkl_2,\frkl_3)$  be  split Lie $2$-algebroids with the structure functions $\mu$   and $\gamma$ respectively. The pair $(\huaA,\huaA^*[1])$ is called a {\bf split Lie $2$-bialgebroid} if $\gamma_2=\mu_2$ and
\begin{equation}
\Poisson{\mu+\gamma-\mu_2,\mu+\gamma-\mu_2}=0.
\end{equation}
\end{defi}
Denote a split Lie $2$-bialgebroid by $(\huaA,\huaA^*[1])$. There is a close relationship between split Lie $2$-bialgebroids and $\LWX$ 2-algebroids. Assume that $(A_{-1},A_0,a, l_1,l_2,l_3)$   and $(A_0^*[1],A_{-1}^*[1],\frka,\frkl_1,\frkl_2,\frkl_3)$ are Lie $2$-algebroids.
Let $E_0=A_0\oplus A^*_{-1}$, $E_{-1}=A_{-1}\oplus A^*_{0}$ and $\huaE=E_0\oplus E_{-1}$.

Let $\partial:E_{-1}\longrightarrow E_0$ and $\rho:E_0\longrightarrow TM$ be bundle maps defined by
\begin{eqnarray}
\label{eq:parbi}\partial(X^1+\alpha^0)&=&l_1(X^1)+\frkl_1(\alpha^0),\label{eq:L2Bpartial}\\
\rho(X^0+\alpha^1)&=&a(X^0)+\frka(\alpha^1)\label{eq:L2Banchor}.
\end{eqnarray}

On $\Gamma(\huaE)$, define the  operation $\diamond$ by
\begin{equation}\label{eq:L2Bbracket0}\left\{\begin{array}{rcl}
(X^0+\alpha^1)\diamond(Y^0+\beta^1)&=&l_2(X^0,Y^0)+L^0_{X^0}\beta^1-L^0_{Y^0}\alpha^1 +\frkl_2(\alpha^1,\beta^1)+\huaL^0_{\alpha^1}Y^0-\huaL^0_{\beta^1}X^0,\\
(X^0+\alpha^1)\diamond(X^1+\alpha^0)&=&l_2(X^0,X^1)+L^0_{X^0}\alpha^0+\iota_{X^1}\delta(\alpha^1)\\
&& +\frkl_2(\alpha^1,\alpha^0)+\huaL^0_{\alpha^1}X^1+\iota_{\alpha^0}\delta_*(X^0), \\
(X^1+\alpha^0)\diamond(X^0+\alpha^1)&=&l_2(X^1,X^0)+L^1_{X^1}\alpha^1-\iota_{X^0}\delta(\alpha^0)\\&& +\frkl_2(\alpha^0,\alpha^1)+\huaL^1_{\alpha^0}X^0-\iota_{\alpha^1}\delta_*(X^1).
\end{array}\right.
\end{equation}

An $E_{-1}$-valued $3$-form $\Omega$ on $E_0$ is defined by
\begin{eqnarray}\label{eq:L2B3-form}
\nonumber\Omega(X^0+\alpha^1,Y^0+\beta^1,Z^0+\zeta^1)
&=&l_3(X^0,Y^0,Z^0)+L^3_{X^0,Y^0}\zeta^1+L^3_{Y^0,Z^0}\alpha^1+L^3_{Z^0,X^0}\beta^1\\
&&+\frkl_3(\alpha^1,\beta^1,\zeta^1)+\huaL^3_{\alpha^1,\beta^1}Z^0+\huaL^3_{\beta^1,\zeta^1}X^0+\huaL^3_{\zeta^1,\alpha^1}Y^0,
\end{eqnarray}
where $X^0,Y^0,Z^0\in\Gamma(A_0),~\alpha^1,\beta^1,\zeta^1\in\Gamma(A_{-1}^*).$ Here $L^0,L^1,L^3$ are the Lie derivatives for the Lie 2-algebroid $\huaA$ given by \eqref{eq:L} and  $\huaL^0,\huaL^1,\huaL^3$ are the Lie derivatives for the Lie 2-algebroid $\huaA^*[1]$.

\begin{thm}{\rm(\cite{LiuSheng})}\label{thm:Lie2biC2}
Let $(\huaA,\huaA^*[1])$ be a split Lie $2$-bialgebroid. Then $(E_{-1},E_0,\partial,\rho,(\cdot,\cdot)_+,\diamond,\Omega)$ is a $\LWX$ $2$-algebroid, where  $E_0=A_0\oplus A^*_{-1}$, $E_{-1}=A_{-1}\oplus A^*_{0}$, $\partial$ is given by \eqref{eq:L2Bpartial}, $\rho$ is given by \eqref{eq:L2Banchor}, $(\cdot,\cdot)_+$ is given by \eqref{eq:naturalsymform}, $\diamond$ is given by \eqref{eq:L2Bbracket0} and $\Omega$ is given by \eqref{eq:L2B3-form}.
\end{thm}

\begin{defi}
  The following two $\VB$-Lie $2$-algebroids
  \begin{equation}\label{eq:VBLie2}
\xymatrix@!0{  D_{-1} \ar@{->} [rr] \ar@ {->} [dd] \ar@{->} [rd]
&&  B  \ar@ {->} [dd] \ar@{->} [rd]&&&\\
&A_{-1}\ar@{->} [rr]\ar@{->} [dd]&&M\ar@{->} [dd]&&C_{-1}\ar@{->}[ll]\\
D_0\ar@{->} [rr] \ar@{->} [rd]&&B \ar@{->} [rd]& &&\\
&A_0\ar@{->} [rr]&&M&& C_0,\ar@{->}[ll]} \qquad
 \xymatrix@!0{  D_0^{\star B} \ar@{->} [rr] \ar@ {->} [dd] \ar@{->} [rd]
&&  B  \ar@ {->} [dd] \ar@{->} [rd]&&&\\
&C_0^*\ar@{->} [rr]\ar@{->} [dd]&&M\ar@{->} [dd]&&A_0^*\ar@{->}[ll]\\
D_{-1}^{\star B}\ar@{->} [rr] \ar@{->} [rd]&&B \ar@{->} [rd]& &&\\
&C_{-1}^*\ar@{->} [rr]&&M&& A_{-1}^*\ar@{->}[ll]}
\end{equation}
constitute a {\bf $\VB$-Lie $2$-bialgebroid} if the underlying Lie $2$-algebroids $(D_{-1}^B, D_0^B,a,l_1,l_2,l_3)$ and $(D_0^{\star B},D_{-1}^{\star B},\frka,\frkl_1,\frkl_2,\frkl_3)$ constitute a Lie $2$-bialgebroid.
\end{defi}

 The double of a $\VB$-Lie bialgebroid is a $\VB$-Courant algebroid. The following result is its higher analogue.

\begin{thm}
  Given a $\VB$-Lie $2$-bialgebroid as in \eqref{eq:VBLie2}, we have a $\VB$-$\LWX$ $2$-algebroid:
   \begin{equation*}
\xymatrix{  D_{-1}\oplus D_0^{\star B} \ar@{->} [rr] \ar@ {->} [dd] \ar@{->} [rd]
&&  B  \ar@ {->} [dd] \ar@{->} [rd]&&\\
&A_{-1}\oplus C_0^*\ar@{->} [rr]\ar@{->} [dd]&&M\ar@{->} [dd]&C_{-1}\oplus A_{0}^*\ar@{->}[l]\\
D_0\oplus D_{-1}^{\star B}\ar@{->} [rr] \ar@{->} [rd]&&B \ar@{->} [rd]& &\\
&A_0\oplus C_{-1}^*\ar@{->} [rr]&&M& C_0\oplus A_{-1}^*.\ar@{->}[l]}
\end{equation*}
\end{thm}
\pf By Theorem \ref{thm:Lie2biC2}, $(D_{-1}^B\oplus D_0^{\star B},D_0^B\oplus D_{-1}^{\star B},\partial,\rho,S,\diamond,\Omega)$ is a $\LWX$ $2$-algebroid, where $\partial$ is given by \eqref{eq:L2Bpartial}, $\rho$ is given by \eqref{eq:L2Banchor}, $(\cdot,\cdot)_+$ is given by \eqref{eq:naturalsymform}, $\diamond$ is given by \eqref{eq:L2Bbracket0} and $\Omega$ is given by \eqref{eq:L2B3-form}. Then it is not hard to see that the structure maps $\partial,~\rho,~S,~\diamond,~\Omega$ are all linear. Thus, the double of a $\VB$-Lie $2$-bialgebroid is a $\VB$-$\LWX$ $2$-algebroid. \qed

\section{$E$-$\LWX$ 2-algebroid}

In this section, we  introduce the notion of an $E$-$\LWX$ $2$-algebroid as the categorification of an $E$-Courant algebroid introduced in \cite{CLS2}. We show that associated to a $\VB$-$\LWX$ $2$-algebroid, there is an $E$-$\LWX$ $2$-algebroid structure on the corresponding graded fat bundle.

 There is an $E$-valued pairing $\ppE{\cdot,\cdot}_E$ between the jet bundle  $\jet{E}$ and the first order covariant differential operator bundle $\dev{E}$
defined by
$$
\ppE{\mu,\frkd}_E~ \defbe
\frkd(u),\quad\forall~ ~ \frkd\in(\dev{E})_m,~\mu\in (\jet{E})_m,~ u\in \Gamma(E)~~\mbox{{\rm statisfying}}~~\mu=[u]_m.
$$
  \begin{defi}\label{defi: $E$-C A}
Let $E$ be a vector bundle. An $E$-$\LWX$ $2$-algebroid is a $6$-tuple $(\huaK,\partial,\rho,\huaS,
\diamond,\Omega)$, where $\huaK=K_{-1}\oplus K_0$ is a graded vector bundle  over $M$ and
\begin{enumerate}
 \item[$\bullet$] $\partial:K_{-1}\longrightarrow K_0$ is a bundle map;
 \item[$\bullet$] $\huaS:\huaK\otimes\huaK\lon E$ is a surjective graded symmetric nondegenerate  $E$-valued
pairing of degree $1$, which induces an embedding: $\huaK \hookrightarrow
\Hom(\huaK,E)$;
 \item[$\bullet$] $\rho:K_0\lon \dev E$ is a bundle map, called the anchor, such that $\rhowx( \jet E)\subset K_{-1}$,
 i.e. $$\huaS(\rhowx(\mu),~e^0)= \ppE{\mu,\rho(e^0)}_E,~ ~~\forall~ \mu\in\Gamma(\jet E),~e^0\in\Gamma(K_0);$$
  \item[$\bullet$]$\diamond:\Gamma(K_{-i})\times \Gamma(K_{-j})\longrightarrow \Gamma(K_{-(i+j)}),~0\leq i+j\leq 1$ is an $\mathbb R$-bilinear operation;
  \item[$\bullet$]$\Omega:\wedge^3 K_0\longrightarrow K_{-1}$ is a bundle map,

\end{enumerate}
such that the following
properties hold:
\begin{itemize}
\item[\rm (E1)]  $(\Gamma(\huaK),\partial,\diamond,\Omega)$ is a Leibniz $2$-algebra;
\item[ \rm (E2)]  for all $e\in\Gamma(\huaK)$, $e\diamond e=\frac{1}{2}\huaD \huaS(e,e)$, where $\huaD:\Gamma(E)\longrightarrow \Gamma(K_{-1})$ is defined by
 \begin{equation}
 \huaS(\huaD u,e^0)=\rho(e^0)(u),\quad \forall u\in\Gamma(E),~e^0\in\Gamma(K_0);
 \end{equation}
\item[\rm (E3)] for all $e^1_1,e^1_2\in\Gamma(K_{-1})$, $\huaS( \partial(e^1_1),e^1_2)=\huaS(e^1_1,\partial(e^1_2))$;
\item[\rm (E4)]for all $e_1,e_2,e_3\in\Gamma(\huaK)$, $\rho(e_1)\huaS( e_2,e_3)=\huaS( e_1\diamond e_2,e_3)+\huaS(e_2,e_1\diamond e_3)$;
 \item[\rm (E5)]  for all $e^0_1,e^0_2,e^0_3,e^0_4\in\Gamma(K_0)$, $\huaS(\Omega(e^0_1,e^0_2,e^0_3),e^0_4)=-\huaS(e^0_3,\Omega(e^0_1,e^0_2,e^0_4))$;
\item[\rm (E6)] for all $e^0_1,e^0_2\in\Gamma(K_0)$, $ \rho(e^0_1\diamond e^0_2) =[\rho(e^0_1),\rho(e^0_2)]_{\dev}$, where $[\cdot,\cdot]_\dev$ is the commutator bracket on $\Gamma(\dev E)$.
 \end{itemize}
\end{defi}
 A $\LWX$ $2$-algebroid can give rise to a Lie 3-algebra (\cite[Theorem 3.11]{LiuSheng}). Similarly, an $E$-$\LWX$ $2$-algebroid can also give rise to a Lie 3-algebra. Consider the graded vector space $\frke=\frke_{-2}\oplus \frke_{-1}\oplus \frke_0$, where $\frke_{-2}=\Gamma(E)$, $\frke_{-1}=\Gamma(K_{-1})$ and $\frke_0=\Gamma(K_0)$. We introduce a skew-symmetric bracket on $\Gamma(\huaK)$,
\begin{equation}\label{eq:skewbracket}
\Courant{e_1,e_2}=\half(e_1\diamond e_2-e_2\diamond e_1),\quad \forall~e_1,e_2\in\Gamma(\huaK),
\end{equation}
which is the skew-symmetrization of $\diamond$.

\begin{thm}\label{thm:ELWX-Lie3}
An $E$-$\LWX$ $2$-algebroid $(\huaK,\partial,\rho,
\huaS,\diamond,\Omega)$ gives rise to a Lie $3$-algebra $(\frke,\frkl_1,\frkl_2,\frkl_3,\frkl_4)$, where $\frkl_i$ are given by
$$\begin{array}{rcll}
\frkl_1(u)&=&\huaD(u),& \forall~ u\in \Gamma(E),\\
\frkl_1(e^1)&=&\partial (e^1),& \forall~ e^1\in \Gamma(K_{-1}),\\
\frkl_2(e^0_1, e^0_2)&=&\Courant{e^0_1,e^0_2},& \forall~ e^0_1, e^0_2\in\Gamma(K_0),\\
\frkl_2(e^0, e^1)&=&\Courant{e^0,e^1},& \forall~ e^0  \in\Gamma(K_0), e^1\in\Gamma(K_{-1}), \\
\frkl_2(e^0, f)&=&\half \huaS(e^0,\huaD f),& \forall~ e^0  \in\Gamma(K_0), f\in \Gamma(E),\\
\frkl_2(e^1_1, e^1_2)&=&0,& \forall~  e^1_1, e^1_2\in\Gamma(K_{-1}),\\
\frkl_3(e^0_1, e^0_2, e^0_3)&=&\Omega(e^0_1, e^0_2, e^0_3),& \forall~ e^0_1, e^0_2, e^0_3\in\Gamma(K_0),\\
 \frkl_3(e^0_1, e^0_2, e^1)&=&-T(e^0_1, e^0_2, e^1),& \forall~ e^0_1, e^0_2\in\Gamma(K_0), e^1\in \Gamma(K_{-1}),\\
\frkl_4(e^0_1, e^0_2, e^0_3, e^0_4)&=&\overline{\Omega}(e^0_1, e^0_2, e^0_3,e^0_4),&\forall~ e^0_1, e^0_2, e^0_3, e^0_4\in\Gamma(K_0),
\end{array}
$$
where the totally skew-symmetric  $T:\Gamma(K_0)\times \Gamma(K_0)\times \Gamma(K_{-1})\longrightarrow \Gamma(E)$ is given by
\begin{equation}
T(e^0_1,e^0_2,e^1)=\frac{1}{6}\big(\huaS(e^0_1,\Courant{e^0_2,e^1})+\huaS(e^1,\Courant{e^0_1,e^0_2})+\huaS(e^0_2,\Courant{e^1,e^0_1})\big),
\end{equation}
and $\overline{\Omega}:\wedge^4\Gamma(K_0) \longrightarrow \Gamma(E)$ is given by
$$
\overline{\Omega}(e^0_1, e^0_2, e^0_3,e^0_4)=\huaS(\Omega(e^0_1, e^0_2, e^0_3),e^0_4).
$$
\end{thm}
\pf The proof is totally parallel to the proof of \cite[Theorem 3.11]{LiuSheng}, we omit the details. \qed\vspace{3mm}

Let $(D_{-1}^B,D_0^B,\partial,\rho,S,\diamond,\Omega)$ be a $\VB$-$\LWX$ $2$-algebroid on the graded double vector bundle $\left(\begin{array}{ccc}
D_{-1};&A_{-1},B;&M\\
D_{0};&A_{0},B;&M
\end{array}\right).$ Then we have the associated graded fat bundles $\hat{A}_{-1}\oplus \hat{A}_0$, which fit the exact sequences:
\begin{eqnarray*}
 &0\rightarrow B^*\otimes A_{0}^* \longrightarrow \hat{A}_{-1}\longrightarrow A_{-1} \rightarrow 0,&\\
 &0\rightarrow B^*\otimes A_{-1}^* \longrightarrow \hat{A}_{0}\longrightarrow A_{0} \rightarrow 0.&
\end{eqnarray*}
Since the bundle map $\partial$ is linear, it induces a bundle map $\hat{\partial}:\hat{A}_{-1}\longrightarrow \hat{A}_{0}$. Since the anchor $\rho$ is linear, it induces a bundle map $\hat{\rho}:\hat{A}_{0}\longrightarrow\dev B^*$, where sections of $\dev B^*$ are viewed as linear vector fields on $B$. Furthermore, the restriction of $S$ on linear sections will give rise to linear functions on $B$. Thus, we obtain a $B^*$-valued degree 1 graded symmetric bilinear form $\hat{S}$ on the graded fat bundle $\hat{A}_{-1}\oplus \hat{A}_0$. Since the operation $\diamond$ is linear, it induces an operation $\hat{\diamond}:\hat{A}_{-i}\times \hat{A}_{-j}\longrightarrow \hat{A}_{-(i+j)}$, $0\leq i+j\leq 1$. Finally, since $\Omega$ is linear, it induces an $\hat{\Omega}:\Gamma(\wedge^3\hat{A_0})\longrightarrow \hat{A}_{-1}$. Then we obtain:

\begin{thm}\label{thm:ELWX2}
  A $\VB$-$\LWX$ $2$-algebroid gives rise to a $B^*$-$\LWX$ $2$-algebroid structure on the corresponding graded fat bundle. More precisely, let $(D_{-1}^B,D_0^B,\partial,\rho,S,\diamond,\Omega)$ be a $\VB$-$\LWX$ $2$-algebroid on the graded double vector bundle $\left(\begin{array}{ccc}
D_{-1};&A_{-1},B;&M\\
D_{0};&A_{0},B;&M
\end{array}\right)$ with the associated graded fat bundle $\hat{A}_{-1}\oplus \hat{A}_0$. Then $(\hat{A}_{-1}, \hat{A}_0,\hat{\partial},\hat{\rho},\hat{S},\hat{\diamond},\hat{\Omega})$ is a $B^*$-$\LWX$ $2$-algebroid.
\end{thm}
\pf Since all the structures defined on the graded fat bundle $\hat{A}_{-1}\oplus \hat{A}_0$ are the restriction of the structures in the $\VB$-$\LWX$ $2$-algebroid, it is straightforward to see that all the axioms in Definition \ref{defi: $E$-C A} hold.\qed

\begin{ex} {\rm
Consider the $\VB$-$\LWX$ $2$-algebroid given in Example \ref{ex:TE}, the corresponding $E$-$\LWX$ $2$-algebroid is
$((\jet E)[1],\dev E,\partial=0,\rho={\id},\huaS=\ppairingE{\cdot,\cdot},\diamond,\Omega=0)$, where the graded symmetric nondegenerate $E$-valued pairing $\ppairingE{\cdot,\cdot}$ is given by
$$
\ppairingE{\frkd+\mu,\frkt+\nu}=\ppE{\mu,\frkt}_E+\ppE{\nu,\frkd}_E,\quad\forall~\frkd+\mu,~\frkt+\nu\in\dev
E\oplus\jet E,
$$
and $\diamond$ is given by
$$
 (\frkd+\mu)\diamond (\frkr+\nu)=
[\frkd,\frkr]_{\dev}+\Lied_{\frkd}\nu-\Lied_{\frkr}\mu +
\jetd\ppE{\mu,\frkr}_E.
$$
See \cite{CLomni} for more details.
 }
\end{ex}

\begin{ex}{\rm
 Consider the $\VB$-$\LWX$ $2$-algebroid given in Proposition \ref{pro:TLWX2}. The graded fat bundle is $\jet E_{-1}\oplus \jet E_0$. It follows that the graded jet bundle associated to a $\LWX$ $2$-algebroid is a $T^*M$-$\LWX$ $2$-algebroid. This is the higher analogue of the result that the jet bundle of a Courant algebroid is $T^*M$-Courant algebroid given in \cite{CLS2}. See also \cite{Jotz} for more details.
 }
\end{ex}

\section{From Lie 3-algebras to Lie 3-algebras}
As applications of $E$-$\LWX$ 2-algebroids introduced in the last section, we construct Lie 3-algebras from Lie 3-algebras in this section. Let $(\g_{-2},\g_{-1},\g_0,l_1,l_2,l_3,l_4)$ be a Lie $3$-algebra. By Theorem \ref{thm:Lie3-VBLWX}, the corresponding $\VB$-$\LWX$ $2$-algebroid  is given by
  \begin{equation*}
\xymatrix{  \g_{-1} \oplus \g_0^* \oplus  \g_{-2}^* \ar@{->} [rr] \ar@ {->} [dd]_{\partial} \ar@{->} [rd]
&& \g_{-2}^* \ar@ {=} [dd] \ar@{->} [rd]&&&\\
&\g_{-1}\ar@{->} [rr]\ar@{->} [dd]_{l_1}&&pt\ar@{=} [dd]&&\g_0^*\ar@{->} [ll]\\
 \g_0 \oplus\g_{-1}^*\oplus \g_{-2}^*\ar'[r] [rr] \ar@{->} [rd]&&\g_{-2}^*\ar@{->} [rd]&&& \\
&\g_0\ar@{->} [rr]&&pt&&\g_{-1}^*.\ar@{->} [ll]}
\end{equation*}

By Theorem \ref{thm:ELWX2}, we obtain:
\begin{pro}
Let $(\g_{-2},\g_{-1},\g_0,l_1,l_2,l_3,l_4)$ be a Lie $3$-algebra. Then there is an $E$-$\LWX$ $2$-algebroid   $(\Hom(\g_0,\g_{-2})\oplus \g_{-1},\Hom(\g_{-1},\g_{-2})\oplus \g_{0},\partial,\rho,\huaS,\diamond,\Omega)$, where for all $x^i, y^i, z^i\in\g_{-i}$, $\phi^i, \psi^i, \varphi^i\in \Hom(\g_{-i},\g_{-2})$, $\partial:\Hom(\g_0,\g_{-2})\oplus \g_{-1}\longrightarrow\Hom(\g_{-1},\g_{-2})\oplus \g_{0}$ is given by
\begin{equation}
  \partial(\phi^0+x^1)=\phi^0\circ l_1+l_2(x^1,\cdot)|_{\g_{-1}}+ l_1(x^1),
\end{equation}
$\rho:\Hom(\g_{-1},\g_{-2})\oplus \g_{0}\longrightarrow\gl(\g_{-2})$ is given by
\begin{equation}\label{eq:Vrho}
  \rho(\phi^1+x^0)=\phi^1\circ l_1+l_2(x^0,\cdot)|_{\g_{-2}},
\end{equation}
the $\g_{-2}$-valued pairing $\huaS$ is given by
\begin{equation}
  \huaS(\phi^1+x^0,\psi^0+y^1)=\phi^1(y^1)+\psi^0(x^0),
\end{equation}
the operation $\diamond$ is given by
\begin{equation}\label{eq:Vcom}\left\{\begin{array}{rcl}
 ( x^0+\psi^1)\diamond (y^0+\phi^1)
    &=&l_2(x^0,y^0)+l_3(x^0,y^0,\cdot)|_{\g_{-1}}+l_2(x^0,\phi^1(\cdot))-\phi^1\circ l_2(x^0,\cdot)|_{\g_{-1}}\\
    &&+\psi^1\circ l_2(y^0,\cdot)|_{\g_{-1}}-l_2(y^0,\psi^1(\cdot))+\psi^1\circ l_1\circ \phi^1-\phi^1\circ l_1\circ \psi^1,\\
  ( x^0+\psi^1 )\diamond (y^1+\phi^0)
     &=&l_2(x^0,y^1)+l_3(x^0,\cdot,y^1)|_{\g_{0}}+l_2(x^0,\phi^0(\cdot))-\phi^0\circ l_2(x^0,\cdot)|_{\g_{0}}\\
     &&-\psi^1l_2(\cdot,y^1)|_{\g_0}+\delta (\psi^1(y^1))+\psi^1\circ l_1\circ \phi^0,\\
 (y^1+\phi^0 )\diamond( x^0+\psi^1 )
     &=&l_2(y^1,x^0)-l_3(x^0,\cdot,y^1)|_{\g_{0}}-l_2(x^0,\phi^0(\cdot))+\phi^0\circ l_2(x^0,\cdot)|_{\g_{0}}\\&& +\delta(\phi^0(x^0))
+\psi^1l_2(\cdot,y^1)|_{\g_0}-\psi^1\circ l_1\circ \phi^0,
\end{array}\right.
\end{equation}
and $\Omega$ is given by
\begin{eqnarray}
  \nonumber &&\Omega(\phi^1+x^0,\psi^1+y^0+\varphi^1+z^0)=l_3(x^0,y^0,z^0)+l_4(x^0,y^0,z^0,\cdot)\\&& \label{eq:Vo}\qquad-\varphi^1\circ l_3(x^0,y^0,\cdot)|_{\g_0}-\phi^1\circ l_3(z^0,x^0,\cdot)|_{\g_0}-\psi^1\circ l_3(y^0,z^0,\cdot)|_{\g_0}.
\end{eqnarray}
\end{pro}

By \eqref{eq:Vrho}, it is straightforward to deduce that the corresponding $\huaD:\g_{-2}\longrightarrow \Hom(\g_0,\g_{-2})\oplus \g_{-1}$ is given by
\begin{equation}\label{eq:VhuaD}
  \huaD(x^2)=l_2(\cdot,x^2)+l_1(x^2)
\end{equation}

Then by Theorem \ref{thm:ELWX-Lie3}, we obtain:
\begin{pro}\label{pro:Lie3}
  Let $(\g_{-2},\g_{-1},\g_0,l_1,l_2,l_3,l_4)$ be a Lie $3$-algebra. Then there is a Lie $3$-algebra
  $(\overline{\g}_{-2},\overline{\g}_{-1},\overline{\g}_0,\frkl_1,\frkl_2,\frkl_3,\frkl_4)$, where $\overline{\g}_{-2}=\g_{-2}$, $\overline{\g}_{-1}=\Hom(\g_0,\g_{-2})\oplus \g_{-1}$, $\overline{\g}_0=\Hom(\g_{-1},\g_{-2})\oplus \g_{0}$, and $\frkl_i$ are given by
  $$\begin{array}{rcll}
\frkl_1(x^2)&=&\huaD(x^2),& \forall~ x^2\in \g_{-2},\\
\frkl_1(\phi^0+x^1)&=&\phi^0\circ l_1+l_2(x^1,\cdot)|_{\g_{-1}}+ l_1(x^1),& \forall~ \phi^0+x^1\in \overline{\g}_{-1},\\
\frkl_2(e^0_1, e^0_2)&=&e^0_1\diamond e^0_2,& \forall~ e^0_1, e^0_2\in\overline{\g}_{0},\\
\frkl_2(e^0, e^1)&=&\half (e^0\diamond e^1-e^1\diamond e^0),& \forall~ e^0  \in\overline{\g}_{0}, e^1\in\overline{\g}_{-1}, \\
\frkl_2(e^0, x^2)&=&\half \huaS(e^0,\huaD x^2),& \forall~ e^0  \in\overline{\g}_{0}, x^2\in \g_{-2},\\
\frkl_2(e^1_1, e^1_2)&=&0,& \forall~  e^1_1, e^1_2\in\overline{\g}_{-1},\\
\frkl_3(e^0_1, e^0_2, e^0_3)&=&\Omega(e^0_1, e^0_2, e^0_3),& \forall~ e^0_1, e^0_2, e^0_3\in\overline{\g}_{0},\\
 \frkl_3(e^0_1, e^0_2, e^1)&=&-T(e^0_1, e^0_2, e^1),& \forall~ e^0_1, e^0_2\in\overline{\g}_{0}, e^1\in \overline{\g}_{-1},\\
\frkl_4(e^0_1, e^0_2, e^0_3, e^0_4)&=&\overline{\Omega}(e^0_1, e^0_2, e^0_3,e^0_4),&\forall~ e^0_1, e^0_2, e^0_3, e^0_4\in\overline{\g}_{0},
\end{array}
$$
where the operation $\huaD$, $\diamond$, $\Omega$ are given by \eqref{eq:VhuaD}, \eqref{eq:Vcom}, \eqref{eq:Vo} respectively, $T:\overline{\g}_0\times\overline{\g}_0\times\overline{\g}_{-1}\longrightarrow \g_{-2}$ is given by
$$
T(e^0_1,e^0_2,e^1)=\frac{1}{6}\big(\huaS(e^0_1,\frkl_2(e^0_2,e^1))+\huaS(e^1,\frkl_2(e^0_1,e^0_2))+\huaS(e^0_2,\frkl_2(e^1,e^0_1))\big),
$$
and $\overline{\Omega}:\wedge^4\overline{\g}_{0} \longrightarrow \g_{-2}$ is given by
$$
\overline{\Omega}(e^0_1, e^0_2, e^0_3,e^0_4)=\huaS(\Omega(e^0_1, e^0_2, e^0_3),e^0_4).
$$
\end{pro}

\emptycomment{
These two Lie 3-algebras are not equivalent, but we have
\begin{thm}
  Let   $(\g_{-2},\g_{-1},\g_0,l_1,l_2,l_3,l_4)$ be a Lie $3$-algebra, and
  $(\overline{\g}_{-2},\overline{\g}_{-1},\overline{\g}_0,\frkl_1,\frkl_2,\frkl_3,\frkl_4)$ is the Lie $3$-algebra given in Proposition \ref{pro:Lie3}. Then there is an $L_\infty$-morphism $(F_0,F_1,F_2,F_3)$ from $(\overline{\g}_{-2},\overline{\g}_{-1},\overline{\g}_0,\frkl_1,\frkl_2,\frkl_3,\frkl_4)$ to $(\g_{-2},\g_{-1},\g_0,l_1,l_2,l_3,l_4)$, where $F_i$ are given by
\end{thm}
}

By Proposition \ref{pro:Lie3}, we can give interesting examples of Lie 3-algebras.

\begin{ex}{\rm
  We view a 3-term complex of vector spaces $V_{-2}\stackrel{l_1}{\longrightarrow}V_{-1}\stackrel{l_1}{\longrightarrow}V_0$ as an abelian Lie 3-algebra. By Proposition \ref{pro:Lie3}, we obtain the Lie 3-algebra $$(V_{-2},\Hom(V_0,V_{-2})\oplus V_{-1},\Hom(V_{-1},V_{-2})\oplus V_0,\frkl_1,\frkl_2,\frkl_3,\frkl_4=0),$$ where $\frkl_i, i=1,2,3$ are given by
  $$\begin{array}{rcl}
\frkl_1(x^2)&=&l_1(x^2),\\
\frkl_1(\phi^0+y^1)&=&\phi^0\circ l_1+ l_1(y^1),\\
\frkl_2(\psi^1+x^0,\phi^1+y^0)&=& \psi^1\circ l_1\circ \phi^1-\phi^1\circ l_1\circ \psi^1,\\
\frkl_2(\psi^1+x^0,\phi^0+y^1)&=&\half l_1(\psi^1(y^1)-\phi^0(x^0))+\psi^1\circ l_1\circ \phi^0,\\
\frkl_2(\psi^1+x^0, x^2)&=&\half \psi^1(l_1( x^2)),\\
\frkl_2(\psi^0+x^1,\phi^0+y^1)&=&0,\\
\frkl_3(\psi^1+x^0,\phi^1+y^0,\varphi^1+z^0)&=&0,\\
 \frkl_3(\psi^1+x^0,\phi^1+y^0,\varphi^0+z^1)&=&-\frac{1}{4}\Big(\psi^1\circ l_1\circ \phi^1(z^1)-\phi^1\circ l_1\circ \psi^1(z^1)\\
 &&-\psi^1\circ l_1\circ \varphi^0(y^0)+\phi^1\circ l_1\circ \varphi^0(x^0)\Big),
\end{array}
$$
for all $ x^2\in V_{-2},~\psi^0+x^1,\phi^0+y^1,\varphi^0+z^1\in\Hom(V_0,V_{-2})\oplus V_{-1},  ~  \psi^1+x^0,\phi^1+y^0,\varphi^1+z^0\in\Hom(V_{-1},V_{-2})\oplus V_0.  $
}
\end{ex}

\begin{ex}{\rm \bf (Higher analogue of the Lie 2-algebra of string type )}\\
  {\rm
  A Lie $2$-algebra $(\g_{-1},\g_0,\widetilde{l_1},\widetilde{l_2},\widetilde{l_3} )$ gives rise to a Lie $3$-algebra $(\mathbb R,\g_{-1},\g_0,{l_1},{l_2},{l_3},{l_4}=0)$ naturally, where ${l_i}$, $i=1,2,3$ is given by
  \begin{eqnarray*}
    {l_1}(r)&=&0,\quad l_1(x^1)=\widetilde{l_1}(x^1),\\
l_2(x^0,y^0)&=&\widetilde{l_2}(x^0,y^0),\quad l_2(x^0,y^1)= \widetilde{l_2}(x^0,y^1),\quad  {l_2}(x^0,r)=0,\quad  {l_2}(x^1,y^1)=0,\\
l_3(x^0,y^0,z^0)&=&\widetilde{l_3}(x^0,y^0,z^0),\quad  {l_3}(x^0,y^0,z^1)=0,
  \end{eqnarray*}
  for all $x^0,y^0,z^0\in\g_0$, $x^1,y^1,z^1\in\g_{-1}$, and $r,s\in\mathbb R$. By Proposition \ref{pro:Lie3}, we obtain the Lie 3-algebra $(\mathbb R,\g_{-1}\oplus \g_0^*,\g_0\oplus \g_{-1}^*,\frkl_1,\frkl_2,\frkl_3,\frkl_4)$, where $\frkl_i$, $i=1,2,3,4$ are given by
  \begin{eqnarray*}
    \frkl_1(r)&=&0,\\ \frkl_1(x^1+\alpha^0)&=&l_1(x^1)+l_1^*(\alpha^0),\\
    \frkl_2(x^0+\alpha^1,y^0+\beta^1)&=&l_2(x^0,y^0)+{\ad^0}^*_{x^0}\beta^1-{\ad^0}^*_{y^0}\alpha^1,\\
\frkl_2(x^0+\alpha^1,y^1+\beta^0)&=&l_2(x^0,y^1)+{\ad^0}^*_{x^0}\beta^0-{\ad^1}^*_{y^1}\alpha^1,\\
\frkl_2(x^1+\alpha^0,y^1+\beta^0)&=&0,\\
\frkl_2(x^0+\alpha^1,r)&=&0,\\
\frkl_3(x^0+\alpha^1,y^0+\beta^1,z^0+\zeta^1)&=&l_3(x^0,y^0,z^0)+{\ad^3}^*_{x^0,y^0}\zeta^1+{\ad^3}^*_{y^0,z^0}\alpha^1+{\ad^3}^*_{z^0,x^0}\beta^1,\\
\frkl_3(x^0+\alpha^1,y^0+\beta^1,z^1+\zeta^0)&=& \half\big(\langle\alpha^1,l_2(y^0,z^1)\rangle+\langle\beta^1,l_2(z^1,x^0)\rangle+\langle\zeta^0,l_2 (x^0,y^0)\rangle\big),\\
\frkl_4(x^0+\alpha^1,y^0+\beta^1,z^0+\zeta^1,u^0+\gamma^1)&=&\langle\gamma^1,l_3(x^0,y^0,z^0)\rangle-\langle\zeta^1,l_3(x^0,y^0,u^0)\rangle\\
&&-\langle\alpha^1,l_3(y^0,z^0,u^0)\rangle-\langle\beta^1,l_3(z^0,x^0,u^0)\rangle
  \end{eqnarray*}
  for all $x^0,y^0,z^0,u^0\in\g_{0},$ $x^1,y^1,z^1\in\g_{-1},$ $\alpha^1,\beta^1,\zeta^1,\gamma^1\in\g^*_{-1},$ $\alpha^0,\beta^0\in\g^*_{0}$, where ${\ad^0}^*_{x^0}:\g^*_{-i}\longrightarrow \g^*_{-i}$, ${\ad^1}^*_{x^1}:\g^*_{-1}\longrightarrow \g^*_{0}$ and ${\ad^3}^*_{x^0,y^0}:\g^*_{-1}\longrightarrow\g^*_{0}$ are defined respectively by
\begin{eqnarray*}
\langle{\ad^0}^*_{x^0}\alpha^1,x^1\rangle&=&-\langle\alpha^1,l_2(x^0,x^1)\rangle,\quad
\langle{\ad^0}^*_{x^0}\alpha^0,y^0\rangle=-\langle\alpha^0,l_2(x^0,y^0)\rangle,\\
\langle{\ad^1}^*_{x^1}\alpha^1,y^0\rangle&=&-\langle\alpha^1,l_2(x^1,y^0)\rangle,\quad
\langle{\ad^3}^*_{x^0,y^0}\alpha^1,z^0\rangle=-\langle\alpha^1,l_3(x^0,y^0,z^0)\rangle.
\end{eqnarray*}
}
\end{ex}

\begin{rmk}
  For any Lie algebra $(\h,[\cdot,\cdot]_\h)$, we have the semidirect product Lie algebra $(\h\ltimes_{\ad^*}\h^*,[\cdot,\cdot]_{\ad^*})$, which is a quadratic Lie algebra naturally. Consequently, one can construct the corresponding Lie $2$-algebra $(\mathbb R,\h\ltimes_{\ad^*}\h^*,l_1=0,l_2=[\cdot,\cdot]_{\ad^*},l_3)$, where $l_3$ is given by
   $$
   l_3(x+\alpha,y+\beta,z+\gamma)=\langle\gamma,[x,y]_\h\rangle+\langle\beta,[z,x]_\h\rangle+\langle\alpha,[y,z]_\h\rangle,\quad \forall x,y,z\in\h,\alpha,\beta,\gamma\in\h^*.
   $$
   This Lie $2$-algebra is called the Lie $2$-algebra of string type in \cite{shengzhu1}. On the other hand, associated to a Lie $2$-algebra $(\g_{-1},\g_0,\widetilde{l_1},\widetilde{l_2},\widetilde{l_3} )$, there is a naturally a quadratic Lie $2$-algebra structure on  $\big(\g_{-1}\oplus \g_0^*\big)\oplus\big(\g_0\oplus\g_{-1}^*\big)$ (\cite[Example 4.8]{LiuSheng}). Thus, the Lie $3$-algebra given in the above example can be viewed as the higher analogue of the Lie $2$-algebra of string type.
\end{rmk}

Motivated by the above example, we show that one can obtain a Lie 3-algebra associated to a quadratic Lie 2-algebra in the sequel. This result is the higher analogue of the fact that there is a Lie 2-algebra, called the string Lie 2-algebra, associated to a quadratic Lie algebra.

A {\bf quadratic Lie 2-algebra} is a Lie 2-algebra $(\g_{-1},\g_0,l_1,l_2,l_3)$ equipped with a degree $1$ graded symmetric nondegenerate bilinear form $S$ which induces an isomorphism between $\g_{-1}$ and $\g_0^*$, such that the following invariant conditions hold:
\begin{eqnarray}
  \label{eq:inv1}S(l_1(x^1),y^1)&=& S(l_1(y^1),x^1),\\
   \label{eq:inv2}S(l_2(x^0,y^0),z^1)&=&-S(l_2(x^0,z^1),y^0),\\
   \label{eq:inv3}S(l_3(x^0,y^0,z^0),u^0)&=&-S(l_3(x^0,y^0,u^0),z^0),
\end{eqnarray}
for all  $x^0,y^0,z^0,u^0\in\g_0,$ $x^1,y^1\in\g_{-1}$.

Let $(\g_{-1},\g_0,l_1,l_2,l_3,S)$ be a quadratic Lie 2-algebra. On the 3-term complex of vector spaces $\mathbb R\oplus \g_{-1}\oplus \g_0$, where $\mathbb R$ is of degree $-2$, we define $\frkl_i$, $ i=1,2,3,4$, by
\begin{equation}
\left\{\begin{array}{rclrcl}
  \frkl_1(r)&=&0,  & \frkl_1(x^1)&=&l_1(x^1),\\
  \frkl_2(x^0,y^0)&=&l_2(x^0,y^0),&\frkl_2(x^0,y^1)&=&l_2(x^0,y^1),\\
  \frkl_2(x^0,r)&=&0,& \frkl_2(x^1,y^1)&=&0,\\
  \frkl_3(x^0,y^0,z^0)&=&l_3(x^0,y^0,z^0),&\frkl_3(x^0,y^0,z^1)&=&\half S(z^1,l_2(x^0,y^0)),\\
  \frkl_4(x^0,y^0,z^0,u^0)&=& S(l_3(x^0,y^0,z^0),u^0),&&&
  \end{array}\right.
\end{equation}
for all  $x^0,y^0,z^0,u^0\in\g_0,$ $x^1,y^1,z^1\in\g_{-1}$ and $r\in\mathbb R$.
\begin{thm}
  Let $(\g_{-1},\g_0,l_1,l_2,l_3,S)$ be a quadratic Lie $2$-algebra. Then $(\mathbb R, \g_{-1}, \g_0,\frkl_1,\frkl_2,\frkl_3,\frkl_4)$ is a Lie $3$-algebra, called {\bf the higher analogue of the string Lie $2$-algebra}.
\end{thm}
\pf It follows from direct verification of the coherence conditions for $\frkl_3$ and $\frkl_4$ using the invariant conditions \eqref{eq:inv1}-\eqref{eq:inv3}. We omit details. \qed



\end{document}